\documentclass{aa}
\usepackage{graphicx}
\usepackage{natbib}
\bibpunct{(}{)}{;}{a}{}{,}

\def \kms{km\,s$^{-1}$}

\def \oiiib{[O{\small III}]$\lambda 5007$}

\newcommand{\hb}{\mbox{H$\beta$}}
\newcommand{\hd}{\mbox{H$\delta$}}
\newcommand{\hdf}{\mbox{H$\delta_F$}}
\newcommand{\hda}{\mbox{H$\delta_A$}}
\newcommand{\hg}{\mbox{H$\gamma$}}
\newcommand{\hgf}{\mbox{H$\gamma_F$}}
\newcommand{\hga}{\mbox{H$\gamma_A$}}
\newcommand{\meanFe}{\mbox{$\langle$Fe$\rangle$}}
\newcommand{\mgb}{\mbox{Mg\,$b$}}

\begin{document}
   \title{VLT spectroscopy of NGC\,3115 globular clusters
   \thanks{Based on observations collected at the European Southern
   Observatory, Cerro Paranal, Chile (ESO No. 66.B-0131).}}

   \author{Harald Kuntschner\inst{1}\inst{,}\inst{4}
          \and
          Bodo L. Ziegler\inst{2}\inst{,}\inst{3}
          \and
          R.M. Sharples\inst{4}
          \and
          Guy Worthey\inst{5}
          \and
          Klaus J. Fricke\inst{3}
          }

   \institute{European Southern Observatory, Karl-Schwarzschild-Str. 2,
              85748 Garching bei M\"unchen, Germany\\
              \email{hkuntsch@eso.org}
         \and
              Academy of Sciences, Theaterstr. 7, 37073 G\"ottingen, Germany
         \and
              Universit\"atssternwarte, Geismarlandstrasse
              11, 37083 G\"ottingen, Germany
         \and
              Department of Physics, University of Durham, Durham DH1
              3LE, UK
         \and
              Department of Physics, Washington State University,
              1245 Webster Hall, Pullman, WA 99164-2814, USA
              }

   \date{Received 16.07.2002; accepted ???}

   \abstract{We present results derived from VLT--FORS2 spectra of 24
     different globular clusters associated with the lenticular galaxy
     NGC\,3115. A subsample of 17 globular clusters have sufficiently
     high signal--to--noise to allow precision measurements of
     absorption line-strengths. Comparing these indices to new stellar
     population models by Thomas et al. we determine ages,
     metallicities and element abundance ratios. For the first time
     these stellar population models explicitly take abundance ratio
     biases in the Lick/IDS stellar library into account. Our data are
     also compared with the Lick/IDS observations of Milky Way and
     M\,31 globular clusters. Unpublished higher order Balmer lines
     (H$\gamma_{A,F}$ and H$\delta_{A,F}$) from the Lick/IDS
     observations are given in the Appendix.
     
     Our best age estimates show that the observed clusters which
     sample the bimodal colour distribution of NGC\,3115 are coeval
     within our observational errors (2--3 Gyr). Our best calibrated
     age/metallicity diagnostic diagram (\hb\/ {\em vs}\/ [MgFe])
     indicates an absolute age of 11--12 Gyr consistent with the
     luminosity weighted age for the central part of NGC\,3115. We
     confirm with our accurate line-strength measurements that the
     $(V-I)$ colour is a good metallicity indicator within the probed
     metallicity range ($-1.5 < \mathrm{[Fe/H]} < 0.0$).  The abundance
     ratios for globular clusters in NGC\,3115 give an inhomogeneous
     picture. We find a range from solar to super-solar ratios for both
     blue and red clusters. This is similar to the data for M\,31 while
     the Milky Way seems to harbour clusters which are mainly
     consistent with $[\alpha / \mathrm{Fe}] \simeq 0.3$. From our
     accurate recession velocities we detect, independent of
     metallicity, clear rotation in the sample of globular clusters.
     
     In order to explain the metallicity and abundance ratio pattern,
     particularly the range in abundance ratios for the metal rich
     globular clusters in NGC\,3115, we favour a formation picture with
     more than two distinct formation episodes.
     
     }

   \maketitle
%

%
%
\section{Introduction}
The analysis of globular cluster systems in external galaxies is
starting to fulfil its long-held promise as a probe of the formation of
galaxies. Globular clusters (hereafter GCs) are a fossil record of this
formation process, and provide one of the best tools with which to
investigate the chemical enrichment and star formation history in the
initial stages of galaxy formation \citep[e.g.,][]{ash97}. Much recent
interest has focused on the globular cluster systems of luminous
elliptical galaxies where the combination of metallicities and
kinematics can be used to distinguish between variants of the popular
monolithic collapse and merger models for galaxy formation
\citep[e.g.,][]{forb97,sharp98}.

Lenticular galaxies hold a key position in the Hubble sequence of
morphological types, intermediate between pure spheroidal systems like
luminous ellipticals and disk-dominated spiral galaxies.  Their
formation mechanism is still the subject of considerable debate with
evidence both for \citep{dres97} and against \citep{dres80} their
evolution from star-forming spirals via processes of gas stripping and
exhaustion. A key question is when and how did such processes occur for
S0 galaxies in a wide range of environments from rich clusters to the
field. The globular cluster systems of S0 galaxies can provide
independent constraints on when the major star formation episodes
occurred both in the disk and halo. However, thus far they have been
little studied with only NGC\,1380 \citep{kis97} and NGC\,4594
\citep[the Sombrero galaxy,][]{bri97} having received any detailed
attention and only with photometric methods.

NGC\,3115 is one of the nearest S0 galaxies ($9.7\pm0.4$~Mpc,
$M_B=-20.1$; \citealp{ton01}) and is located in the sparse low-density
environment of the Leo Spur. As such it provides an ideal test case for
studying the formation mechanism of field S0's. A significant globular
cluster system containing $\sim$500 clusters was first detected by
\cite{han86} using photographic plates. The nature of the cluster
system and its origin were recently thrown into question with the
discovery by \cite{els97} that the red-giant stars in the NGC\,3115
halo $\sim$40~kpc from the centre showed a {\em bimodal}\/ colour
distribution. The inferred presence of two distinct halo populations of
roughly equal size at metallicities of $\mathrm{[Fe/H]} \simeq -0.7$
and $\mathrm{[Fe/H]} \simeq -1.3$ suggests at least two distinct epochs
of formation.

The $(V-I)$ colour distribution of the NGC\,3115 globular cluster
system has been the target of two recent independent studies using HST
\citep{kundu98} and CFHT \citep{kav98b} data. Both studies find
bimodality in the colour distributions of the GCs, with mean
metallicities at $\mathrm{[Fe/H]} \simeq -0.37$ and $\mathrm{[Fe/H]}
\simeq -1.36$ suggesting that the cluster and halo star systems may
have formed coevally. This suggestion has gained further support from
an investigation by \citet{puz02a} who employed optical--IR colours to
probe the globular cluster population close to the centre of NGC\,3115.
They also find two peaks in metallicity and an average age around
$\approx$10~Gyr. However, their age discrimination power is very
limited for metallicities lower than ${\rm [Fe/H]} = -0.4$.

One scenario in which the above observations could be understood is if
the metal-poor component corresponds to a primordial $\simeq 13$~Gyr
old population, whilst the metal-rich component formed a few Gyr later
from enriched gas, possibly as the result of a minor merger
\citep[e.g.,][]{bekk98}. With only broad-band colours available,
however, the well-known degeneracy between metallicity and age
\citep{wor94} makes such conclusions very uncertain.

For that reason, we have started a campaign to spectroscopically study
the globular cluster system in NGC\,3115. Our precision measurements of
absorption line-strengths can be used to derive age and metallicity
estimates directly from the comparison with new stellar population
models. Unlike photometric methods, with spectroscopy we are also able
to explore element abundance ratios for the GCs. We compare our results
with other very recently obtained spectroscopic samples of GCs in
early-type galaxies: in the giant Fornax elliptical NGC\,1399
\citep{for01}, the Sa/Sb galaxy M\,81 \citep{schro02}, the SB0 galaxy
NGC\,1023 \citep{lar02a} and the Sombrero galaxy NGC\,4594
\citep{lar02b}.

The paper is organized as follows. In Section~\ref{sec:obs}, the
observations and their reduction are discussed.
Section~\ref{sec:colour} presents the colour distribution of our sample
while in Section~\ref{sec:abrat} the treatment of abundance ratios and
new stellar population models are investigated. Our results on
abundance ratios, age and metallicity distributions for GCs in
NGC\,3115, the Milky Way (hereafter MW) and M\,31 are presented in
Section~\ref{sec:results} with a general discussion in
Section~\ref{sec:discussion}. We present our conclusions in
Section~\ref{sec:conclusions}.

\section{Observations and data reduction}
\label{sec:obs}
%
%
%
%
The candidate GCs were selected from the HST/WFPC2 investigation of
\citet{kundu98} who detected 144 globular cluster candidates in the
central region of \object{NGC\,3115}. In order to keep integration
times reasonably short only clusters with $V<22$ (the peak of the GC
luminosity function is at $V=22.37\pm0.05$; \citealp{kundu98}) were
selected while keeping a balance between red ($(V-I) < 1.06$) and blue
($(V-I) \ge 1.06$) clusters. In order to utilize the full field-of-view
of our multi-object spectrograph we supplemented this list with GC
candidates from a low resolution spectroscopy survey reported in
\citet{kav98b} and also placed some slits on promising objects without
prior information (see Table~\ref{tab:objects}).  In total two masks
with each 26 slitlets covering more than 50 different GC candidates
were manufactured.

The observations were carried out 26/27 Feb 2001 with FORS2 on VLT
using the blue 600 l/mm grism and 1\arcsec\/ wide slitlets giving a
resolution of $\sim$4.8 \AA\/ (FWHM) sampled at 1.2~\AA\,pixel$^{-1}$.
The seeing was generally FWHM$\le 1\farcs0$. The total exposure time
for the first mask was 12\,440~s split up in six individual exposures
of varying length.  Due to bad weather the second mask was only exposed
for 2700~s which was not long enough to be of use for this study. The
flux standard GD\,71 was observed through a long-slit (2\farcs5 width)
to enable us to correct the continuum shape of the spectra.

The standard data reduction procedures (bias subtraction,
flat-fielding, wavelength calibration, sky-subtraction and continuum
correction to a relative flux scale) were performed with a combination
of \texttt{MIDAS} and \texttt{IRAF} tasks. For each slit a
2-dimensional subsection of the CCD was extracted and then treated as a
long-slit observation. The extraction region was defined by tracing the
globular cluster spectrum along the wavelength direction and extracting
the corresponding sections for the flat-field and arc-lamp
observations.  The cosmic rays were removed with the routine {\tt
  lacos\_spec} \citep{vdok01}. The wavelength calibration was accurate
to $<0.2$~\AA. After sky-subtraction GCs were extracted from the
individual exposures to achieve an optimal S/N \citep{hor86}.  Finally,
the spectra of the individual exposures were combined.

We used the spectroscopic flux standard GD\,71 to correct the continuum
shape of our spectra to a relative flux scale and applied a reddening
correction of E$(B-V) = 0.146$ \citep{schle98}. In order to transform
our observations onto the Lick/IDS system we convolved our spectra with
a wavelength-dependent Gaussian kernel (taking into account small
variations of the instrumental resolution with position on the chip)
thereby reproducing the Lick/IDS spectral resolution \citep{worott97}.

As a first analysis step we measured the recession velocity (with {\tt
  fxcor} in {\tt IRAF}) and S/N of the GC candidates. The first slit
(Slitlet ID: 1) did not show any object which we ascribe to a field
distortion towards the edges of the field-of-view. Basic information
for the remaining 28 globular cluster candidates is listed in
Table~\ref{tab:objects}. The first column in this table is an object
identifier, the second column indicates from which source the object
was selected from. The following columns list the J2000.0 coordinates
and $V$, $I$ photometry of \citet{kundu98}. The seventh column shows
pseudo $(V-I)$ colours derived from our spectroscopy (see
Section~\ref{sec:colour} for details). Columns eight and nine list the
radial velocity and the mean S/N per pixel respectively. Finally,
column ten indicates whether the GC candidate is a spectroscopically
confirmed member of the NGC\,3115 system (see next paragraph).

   \begin{table*}
      \caption[]{List of observed globular cluster candidates.}
         \label{tab:objects}
     $$
     \begin{tabular}{clllcccrrc}
            \hline
            \noalign{\smallskip}
            Slitlet ID& Origin$^b$ & \multicolumn{2}{c}{RA J2000.0 DEC} & $V$ (HST) & $(V-I)$ (HST) & $(V-I)$ (spec)$^c$ &cz  & S/N & member \\
            \noalign{\smallskip}
            \hline
            \noalign{\smallskip}
  1 &      - & \multicolumn{7}{c}{No object in slit.} \\
  2 &      - & 10$^h$05$^m$06\farcs5 & -7\degr45$^m$01\arcsec & -               & -                & $1.183$ &  408$\pm$37 &  9 & $\surd$\\
  3 &      - & 10$^h$05$^m$06\farcs1 & -7\degr44$^m$30\arcsec & -               & -                & $1.142$ &  439$\pm$17 & 23 & $\surd$\\
  4 &      - & 10$^h$05$^m$09\farcs9 & -7\degr45$^m$01\arcsec & -               & -                & $0.883$ &    0$\pm$11 & 72 & -\\
  5 & WHT09  & 10$^h$05$^m$11\farcs4 & -7\degr45$^m$01\arcsec & -               & -                & $0.980$ &  379$\pm$21 & 32 & $\surd$\\
  6 &      - & 10$^h$05$^m$09\farcs2 & -7\degr43$^m$52\arcsec & -               & -                & -       &  z=0.39$^a$ & -  & -      \\
  7 & HST08  & 10$^h$05$^m$12\farcs6 & -7\degr44$^m$28\arcsec & $20.282\pm0.006$& $1.178\pm0.008$  & $1.148$ &  291$\pm$16 & 28 & $\surd$\\
  8 & HST03  & 10$^h$05$^m$12\farcs6 & -7\degr44$^m$09\arcsec & $19.794\pm0.006$& $1.051\pm0.009$  & $1.029$ &  393$\pm$13 & 61 & $\surd$\\
  9 & HST02  & 10$^h$05$^m$13\farcs1 & -7\degr43$^m$59\arcsec & $19.742\pm0.005$& $1.077\pm0.007$  & $1.068$ &  494$\pm$16 & 38 & $\surd$\\
 10 & HST21  & 10$^h$05$^m$13\farcs0 & -7\degr43$^m$41\arcsec & $21.072\pm0.012$& $0.979\pm0.018$  & $1.050$ &  456$\pm$37 &  8 & $\surd$\\
 11 & HST39  & 10$^h$05$^m$14\farcs7 & -7\degr43$^m$56\arcsec & $21.539\pm0.014$& $1.199\pm0.018$  & $1.218$ &  702$\pm$31 &  8 & $\surd$\\
 12 & HST13  & 10$^h$05$^m$14\farcs4 & -7\degr43$^m$35\arcsec & $20.544\pm0.008$& $0.979\pm0.011$  & $0.982$ &  582$\pm$35 & 19 & $\surd$\\
 13 & HST18  & 10$^h$05$^m$15\farcs2 & -7\degr43$^m$34\arcsec & $20.970\pm0.010$& $0.989\pm0.013$  & $1.015$ &  546$\pm$28 & 12 & $\surd$\\
14a & HST27  & 10$^h$05$^m$16\farcs6 & -7\degr43$^m$29\arcsec & $21.266\pm0.011$& $1.105\pm0.015$  & $1.117$ &  321$\pm$25 & 11 & $\surd$\\
14b & HST12  & 10$^h$05$^m$17\farcs2 & -7\degr43$^m$20\arcsec & $20.522\pm0.007$& $1.159\pm0.009$  & $1.138$ &  696$\pm$15 & 23 & $\surd$\\
 15 & HST17  & 10$^h$05$^m$16\farcs8 & -7\degr43$^m$09\arcsec & $20.946\pm0.010$& $0.978\pm0.013$  & $1.021$ &  739$\pm$23 & 16 & $\surd$\\
 16 & HST32  & 10$^h$05$^m$18\farcs0 & -7\degr43$^m$15\arcsec & $21.467\pm0.012$& $0.913\pm0.016$  & $0.878$ &  599$\pm$54 & 12 & $\surd$\\
 17 & HST09  & 10$^h$05$^m$17\farcs2 & -7\degr42$^m$49\arcsec & $20.393\pm0.007$& $1.147\pm0.010$  & $1.167$ &  813$\pm$17 & 23 & $\surd$\\
 18 & HST31  & 10$^h$05$^m$18\farcs4 & -7\degr42$^m$45\arcsec & $21.392\pm0.012$& $1.190\pm0.016$  & $1.236$ &  210$\pm$23 &  9 & $\surd$\\
 19 & HST57  & 10$^h$05$^m$17\farcs3 & -7\degr42$^m$08\arcsec & $21.921\pm0.018$& $0.891\pm0.028$  & $0.983$ &  736$\pm$111&  5 & $\surd$\\
 20 & HST46  & 10$^h$05$^m$17\farcs6 & -7\degr41$^m$57\arcsec & $21.802\pm0.015$& $0.919\pm0.023$  & $0.946$ &  661$\pm$55 &  7 & $\surd$\\
 21 &      - & 10$^h$05$^m$20\farcs1 & -7\degr42$^m$16\arcsec & -               & -                & $0.971$ &  846$\pm$42 & 12 & $\surd$\\
 22 &      - & 10$^h$05$^m$19\farcs0 & -7\degr41$^m$47\arcsec & -               & -                & $0.957$ &  670$\pm$41 &  7 & $\surd$\\
 23 &  WHT15 & 10$^h$05$^m$20\farcs2 & -7\degr41$^m$45\arcsec & -               & -                & $1.140$ &  682$\pm$15 & 34 & $\surd$\\
 24 &  WHT16 & 10$^h$05$^m$21\farcs2 & -7\degr41$^m$38\arcsec & -               & -                & $0.992$ & 1008$\pm$18 & 42 & $\surd$\\
25a &      - & 10$^h$05$^m$23\farcs4 & -7\degr41$^m$59\arcsec & -               & -                & $0.832$ &  202$\pm$42 & 19 & $\surd$\\
25b &      - & 10$^h$05$^m$23\farcs7 & -7\degr41$^m$55\arcsec & -               & -                & $0.890$ &  721$\pm$30 & 46 & $\surd$\\
26a &      - & 10$^h$05$^m$22\farcs7 & -7\degr41$^m$13\arcsec & -               & -                & $1.068$ &  803$\pm$24 & 10 & $\surd$\\
26b &  WHT17 & 10$^h$05$^m$22\farcs9 & -7\degr41$^m$07\arcsec & -               & -                & $0.934$ &  769$\pm$25 & 19 & $\surd$\\
            \noalign{\smallskip}
            \hline
         \end{tabular}
     $$
\begin{list}{}{}
\item[$^{\mathrm{a}}$] Background galaxy where [O{\sc ii}] emission is visible.
\item[$^{\mathrm{b}}$] HSTxx denotes globular clusters selected from
  the HST study of \citet{kundu98}, WHTxx are objects from the
  investigation of \citet{kav98b} while ``-'' indicates objects selected
  from our own VLT-FORS2 image in order to fill the mask.
\item[$^{\mathrm{c}}$] Pseudo $(V-I)$ colour derived from the globular
  cluster spectra. The mean error is 0.037~mag. See Section
  \ref{sec:colour} for details.
\end{list}
\end{table*}
   
The radial velocity of NGC\,3115 is listed as $720\pm5$~\kms\/ by
\cite{smi00}. \citet{kav98b} find a mean velocity of 620~\kms\/ and a
velocity dispersion of $\sigma = 177$~\kms\/ from low resolution
spectroscopic observations of 22 globular clusters. We consider all
objects with radial velocities between 200~\kms\/ and 1300~\kms\/ as
members of the NGC\,3115 system. Object~6 is a background galaxy with a
redshift of $z=0.39$, while object~4 is likely to be a galactic star.
In total 26 out of 28 GC candidates (93\%) turned out to be objects
which are dynamically associated with NGC\,3115. This low contamination
fraction is to be expected for a sample of candidates predominantly
selected by morphological criteria from HST data.
   
\begin{figure*}
\centering
\includegraphics[width=15cm]{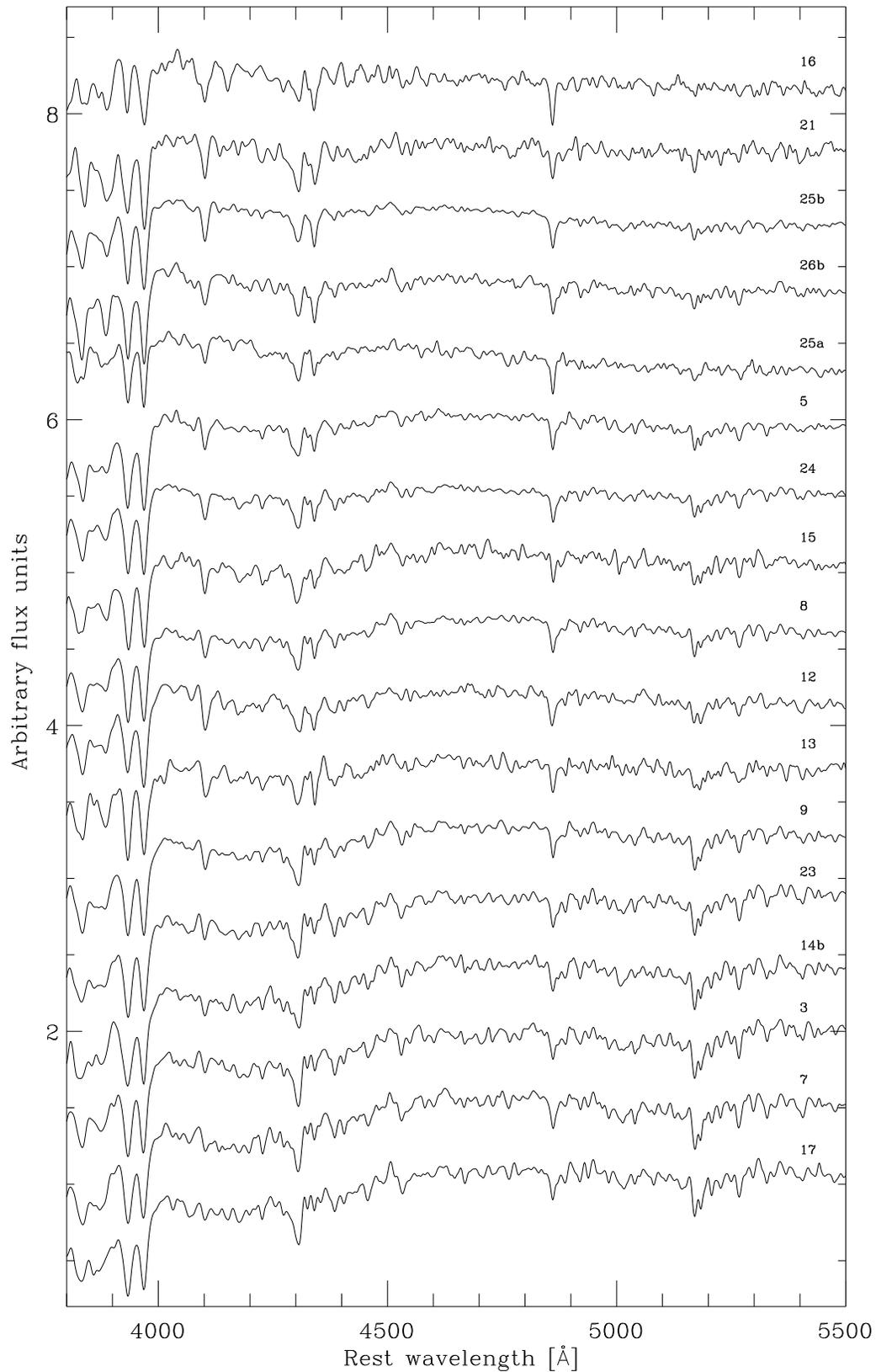}
\caption{Rest-frame spectra of globular clusters corrected
  to the Lick/IDS resolution and sorted by increasing strength of our
  mean metallicity indicator [MgFe] (from top to bottom, see also
  Section~\ref{sec:abrat}). Note the change in overall continuum shape
  (metal poorer globular clusters are bluer) and also the change in
  absorption strength of the Balmer lines (H$\delta$ at 4101~\AA,
  H$\gamma$ at 4340~\AA\/ and H$\beta$ at 4861~\AA). The slitlet ID for
  each spectrum is indicated at the right hand side of the plot (see
  also Table~\ref{tab:objects}).}
           \label{fig:spectra}%
\end{figure*}

In order to allow a proper line-strength analysis we restrict our
sample to GCs with a mean S/N~$\ge 12$ per pixel\footnote{The S/N is
  determined as an average over the observed wavelength range. The
  minimum S/N per resolution element is $\approx$25.} yielding a final
sample of 17 GCs in NGC\,3115. Below this signal-to-noise cut the
line-strength errors are too large to allow a proper age-metallicity
analysis. The mean S/N of our final sample is 27 per pixel (ranging
from 12 to 61). The GCs have a mean velocity of 600~\kms\/ (1$\sigma$
scatter: 215~\kms). The spectra of these 17 GCs broadened to the
Lick/IDS resolution are shown in Fig.~\ref{fig:spectra}.

We note that the globular cluster 14a (HST27, S/N=11) shows clear
\oiiib\/ emission which is however redshifted by $\approx$190~\kms\/
with respect to the recession velocity of the globular cluster itself.
Therefore the emission cannot be associated with the globular cluster
itself, although the spatial extent of the emission is consistent with
the size of the GC on the sky.

Lick indices \citep[for index definitions see][]{worott97,tra98} were
measured from the resolution and continuum corrected spectra covering a
common observed wavelength range of 3670--5500~\AA. Uncertainties in
the indices were derived by Monte-Carlo simulations which take into
account the photon noise, read-out noise of the CCD and the errors in
the velocity determination. The Lick system is based on non-flux
calibrated spectra, so one expects small offsets \citep[e.g.,][]{kun00}
in the zero-point of the line indices. Since no stars in common with
the Lick stellar library were observed during our observing run we use
the offsets established by \citet{vaz99} for a large flux-calibrated
sample of stars (`Jones library') to transform our index measurements
onto the Lick system.  Since typical metallicities for GCs are between
$-2.0 < \mbox{[Fe/H]} < 0.0$, we calculated the offset by averaging the
values for [Fe/H] = $-0.7$ and $-0.4$ given by \citet[][ see his
Table~2]{vaz99}. Note, that some indices (e.g., Mg$_2$) show strong
evidence for a metallicity dependent offset, which can introduce
systematic offsets as a function of line-strength. Here in this paper
we use mainly indices where the former problem is only a second order
effect. The measured index values and their associated errors are
listed in Tables~\ref{tab:ngc3115_lst1} \& \ref{tab:ngc3115_lst2} in
the Appendix.

\begin{figure}
\centering
\resizebox{\hsize}{!}{\includegraphics{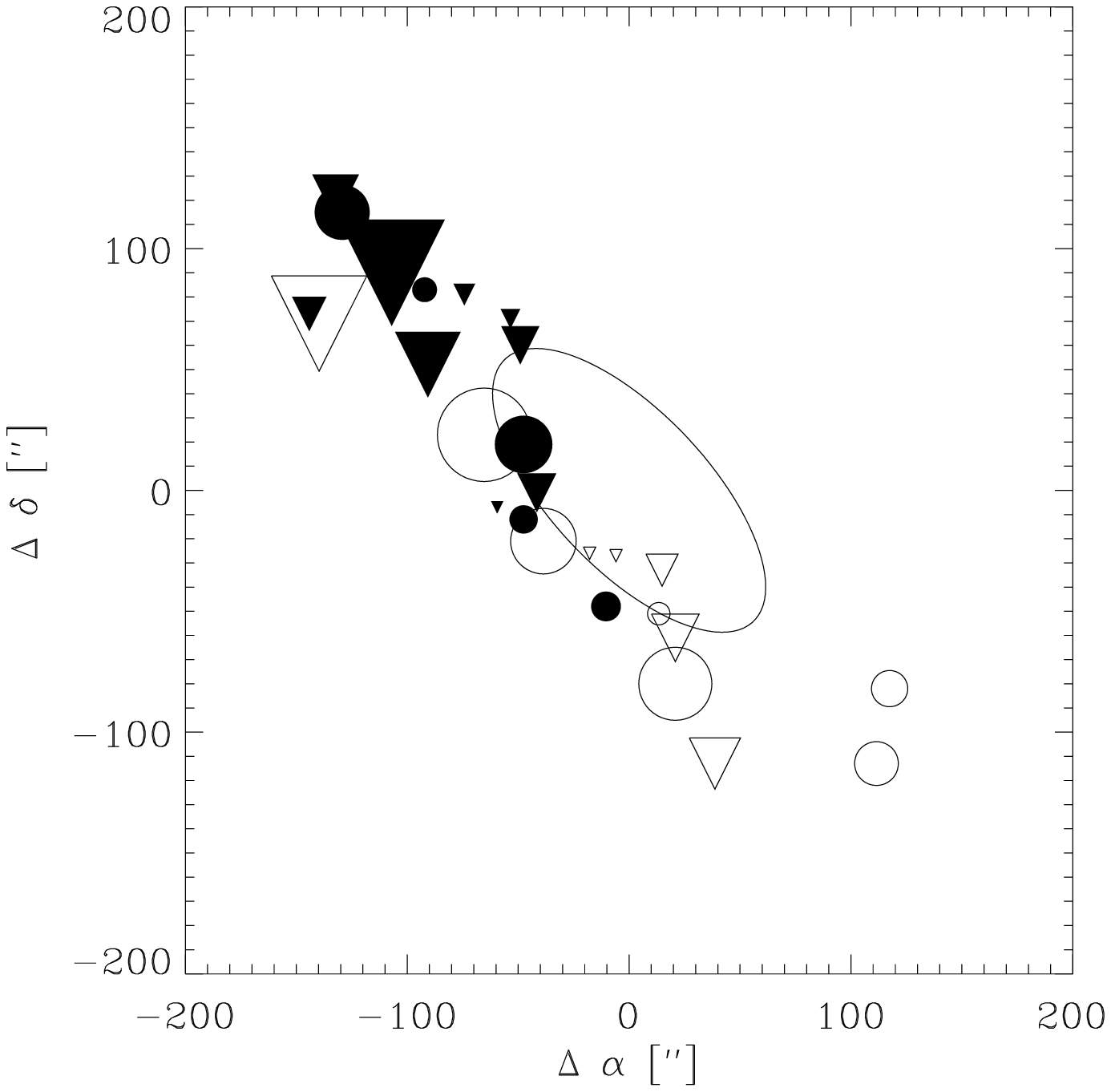}}
\caption{Positions of spectroscopically confirmed globular clusters
  with respect to the centre of NGC\,3115 (triangles represent blue
  clusters and circles red clusters, see Section~\ref{sec:colour}).
  Approaching and receding globular clusters are indicated by open
  symbols and filled symbols respectively, while the symbol size
  indicates the extent of the velocity difference with respect to the
  mean velocity of the whole sample. The ellipse encloses half of the
  integrated light and indicates the position and orientation of the
  main galaxy itself \citep[data from][]{MM94}.  North is up and east
  to the left.}
           \label{fig:pos}%
\end{figure}

The positions of the GCs with respect to the parent galaxy are shown in
Fig.~\ref{fig:pos}. Due to the optimisation of the target efficiency
and the observational constraints of the FORS2 instrument the GCs lie
in a narrow band $\sim$25\arcsec\/ eastwards of the galaxy centre
parallel to the major axis of NGC\,3115. Both the red and blue
sub-samples were evenly distributed across the CCD.

\begin{figure}
\centering
\resizebox{\hsize}{!}{\includegraphics{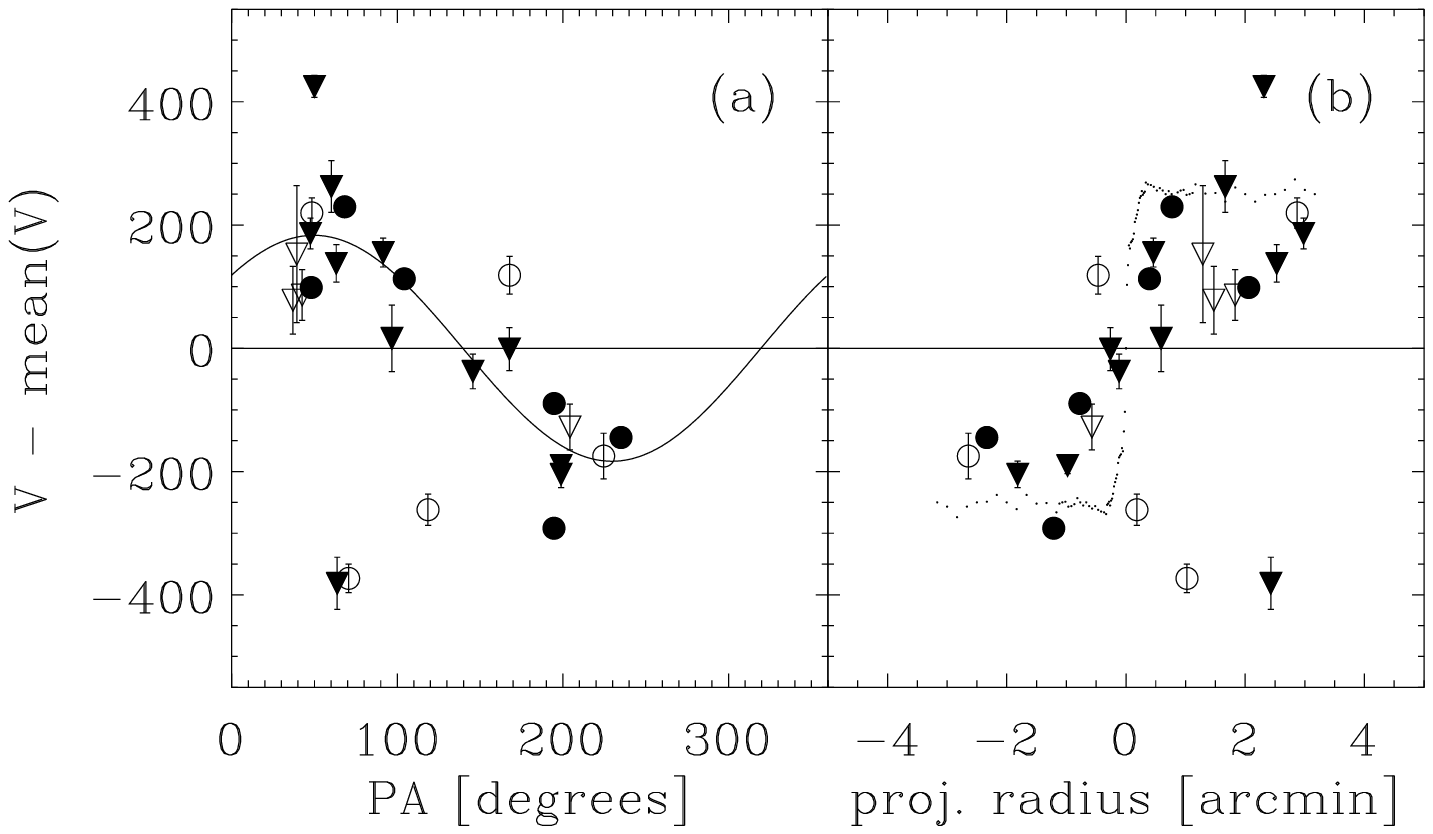}}
\caption{Assessment of rotation for NGC\,3115 globular
  clusters. The triangles and circles represent blue and red clusters,
  respectively.  Filled symbols indicate the high S/N sample of 17
  globular clusters for which we present a line-strength analysis in
  this paper. Panel~a): The vertical axis shows the radial velocity of
  the globular clusters relative to the mean velocity of the sample.
  The horizontal axis is the position angle between a globular cluster
  and the galaxy centre, where a PA of 0 degrees represents north and
  positive is east of north. The sinusoid is a simple least-squares fit
  to the full sample with one iteration to reject outliers. Panel~b):
  radial velocity is plotted against the projected major axis radius.
  Positive numbers are towards north-east. The small dots represent the
  {\em stellar}\/ rotation curve of NGC\,3115 along the major axis
  \citep{cap93}.}
           \label{fig:kin}%
\end{figure}

In Figure~\ref{fig:kin} we assess the kinematics of all globular
clusters which have recession velocities consistent with being members
of the NGC\,3115 system. There is a strong signature of GC rotation
along the major axis of the galaxy. We confirm here the weak signal of
rotation for the red cluster population detected by \citet{kav98b}. In
our sample there is no clear difference between red and blue clusters,
both show an equally strong signal of rotation.
  
Since our sample of GCs is dominated by clusters close to the location
of the major axis (see Figure~\ref{fig:pos}) we note that the sample is
probably biased to globular clusters associated with the disk-formation
of NGC\,3115.

\section{Colour distribution}
\label{sec:colour}
\begin{figure}
\centering
\resizebox{\hsize}{!}{\includegraphics{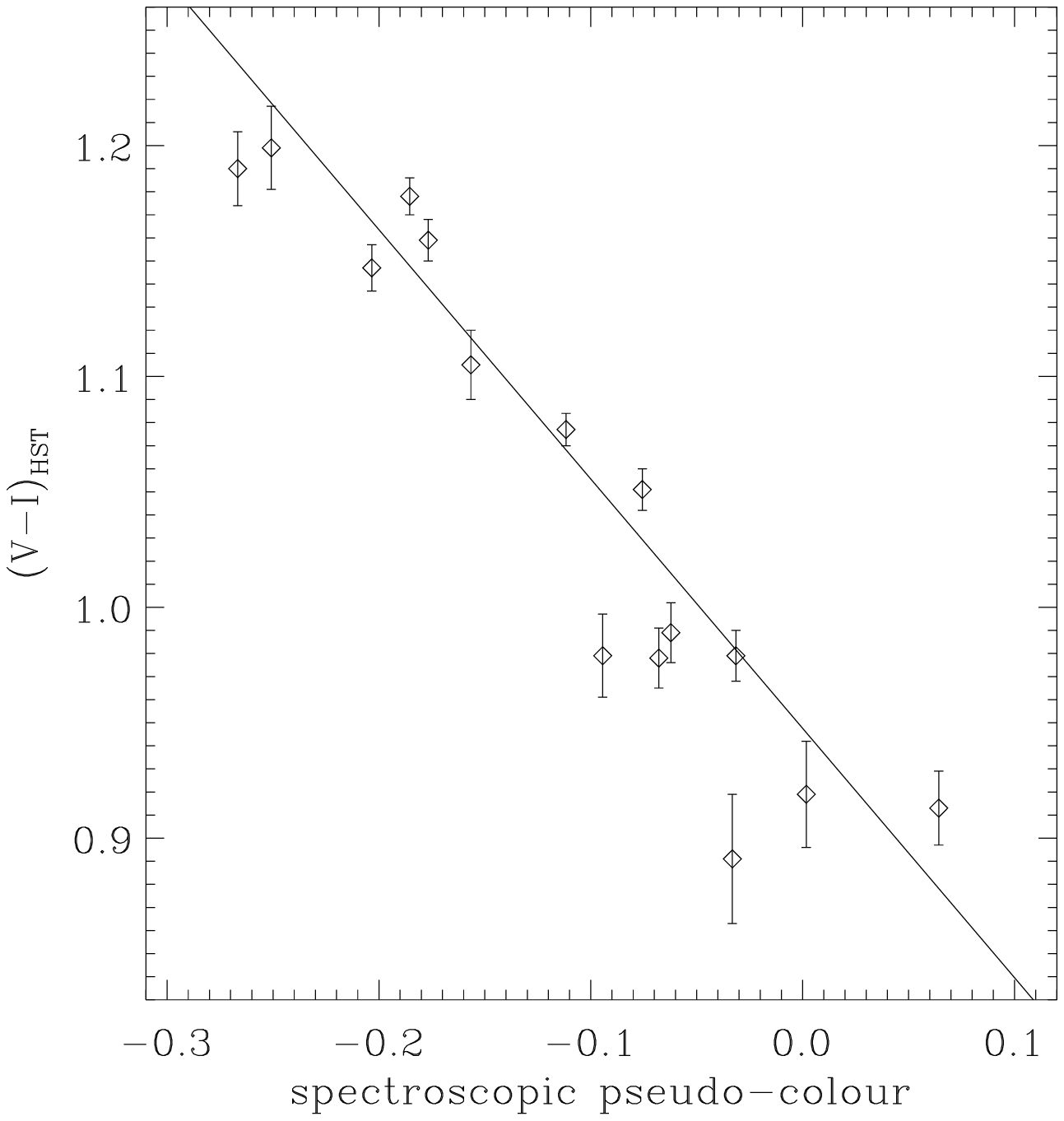}}
\caption{The pseudo colour derived from the 
  spectra versus $(V-I)_{\rm HST}$. The linear fit corresponds to
  $(V-I)_{\rm HST} = -1.08(\pm0.04) \times (spec\; colour) +
  0.95(\pm0.01)$. For details see Section~\ref{sec:colour}.}
           \label{fig:col}%
\end{figure}

9 out of 17 GCs for which we have good quality spectra and are
kinematically associated with NGC\,3115 are within the HST/WFPC2 image.
In order to assign colours to the remaining 8 GCs we derived an
empirical relationship between a ``pseudo-colour'' measured from our
flux-calibrated spectra and the HST $(V-I)$ colours. Of course our
limited wavelength range does not allow us to measure a real $(V-I)$
colour but one can get a rough idea on the overall spectral shape. On
the spectra we measured the mean counts in two regions (3900 --
4600~\AA\/ and 4800 -- 5500~\AA) and plotted the difference versus the
HST $(V-I)$ colour (see Fig.~\ref{fig:col}). This procedure yields a
linear relationship with a mean scatter of 0.037~mag in derived $(V-I)$
colour. Note that here we have used all available spectra of GC
candidates with measured HST colours (see Table~\ref{tab:objects}),
specifically including the ones with an ${\rm S/N} < 12$ per pixel. The
derived $(V-I)$ colours are listed in Table~\ref{tab:objects} in column
seven where we have assigned a mean error of 0.037~mag.

\begin{figure}
\resizebox{\hsize}{!}{\includegraphics{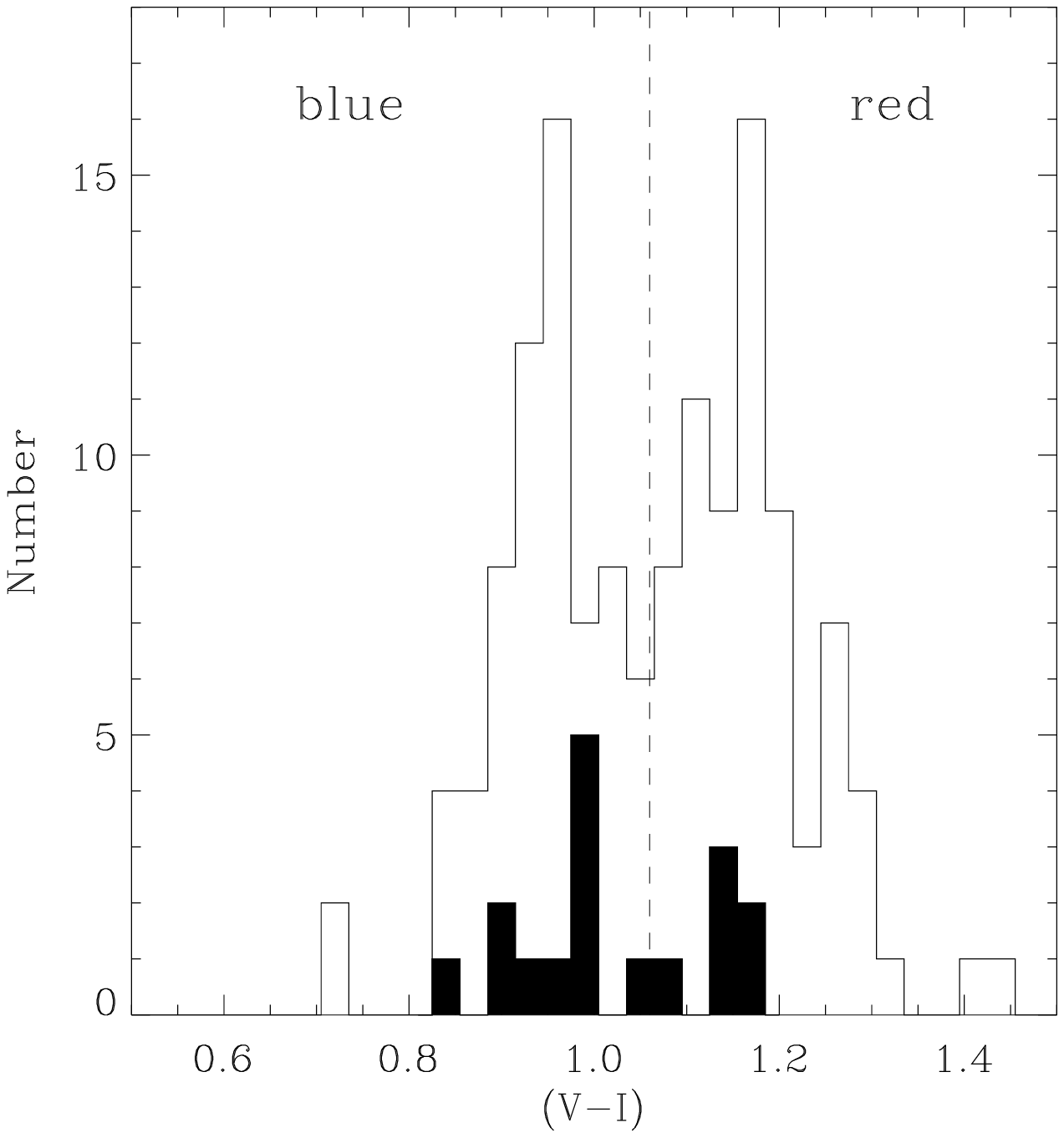}}
\caption{Histogram of $(V-I)$ colours of the cluster candidates from
  \citet{kundu98}. Overplotted as a filled histogram is our
  spectroscopic sample of 17 confirmed globular clusters in NGC\,3115.
  The vertical dashed line indicates the dividing line between red and
  blue clusters at $(V-I) = 1.06$.}
           \label{fig:colors}%
 \end{figure}

The colour distribution of our sample of 17 GCs is shown in
Fig.~\ref{fig:colors} together with the full HST sample by
\citet{kundu98}. The HST sample shows a clear bimodal structure with
peaks at $(V-I) = 0.96$ and $(V-I) = 1.17$. Taking the mean colour of
the HST sample ($(V-I) = 1.06$) as the dividing line between the red
and blue clusters, our spectroscopic sample features eleven blue
clusters and six red ones while spanning the range $(V-I) = 0.87 -
1.18$.

\section{Treatment of abundance ratios}
\label{sec:abrat}
Theory of chemical enrichment predicts that stellar populations created
in a short burst of star formation show elevated magnesium-to-iron
abundance ratios while extended periods of star formation lead to
roughly solar abundance ratios
\citep[e.g.,][]{tin79,WFG92,wor98,TGB99}. This is generally explained
by either a delay in the production of Fe--peak elements with respect
to $\alpha$--elements due to the different timescales of SN\,Ia and
SN\,II, or by a star-burst with an initial mass function skewed to
massive stars. For example, many observations of bright elliptical
galaxies in clusters show indeed elevated magnesium-to-iron ratios
indicating a rapid star formation
\citep[e.g.,][]{pel89,WFG92,DSP93,kun00}.

GCs represent simple stellar populations (hereafter SSP, i.e., a unique
age and chemical composition) since all the stars of one GC are thought
to be created in a single star formation event. But it is not known a
priori whether the stars of a given GC are formed out of Fe-deficient
gas clouds, which have been only enriched by SN\,II producing little
Fe, or out of an already well mixed interstellar medium harbouring the
products from both SN\,Ia (main producer of Fe) and SN\,II. For GC
systems with a bimodal colour distribution like in NGC\,3115, all
scenarios that have been proposed to explain the origin of the red
metal-rich GCs start from the principle that the red population is
formed in a separate star formation event \citep[e.g.,
][]{AZ92,forb97}. In a naive star-formation scenario, where the red
clusters form from the well-mixed interstellar medium they should show
solar abundance ratios.

Most of the currently available stellar population models which can be
used to investigate the abundance ratios of extragalactic objects are
based on stellar libraries from our own Galaxy. This has the
disadvantage that, particularly at sub-solar metallicities, galactic
disk stars show super-solar abundance ratios for many $\alpha$-elements
\cite[][]{edv93,mcw97}. Therefore, without correction, the model
predictions will be biased towards super-solar abundance ratios
\citep[e.g.,][]{BIDT95,kun00,TMB02b}.

\begin{figure*}
\centering \resizebox{\hsize}{!}{\includegraphics{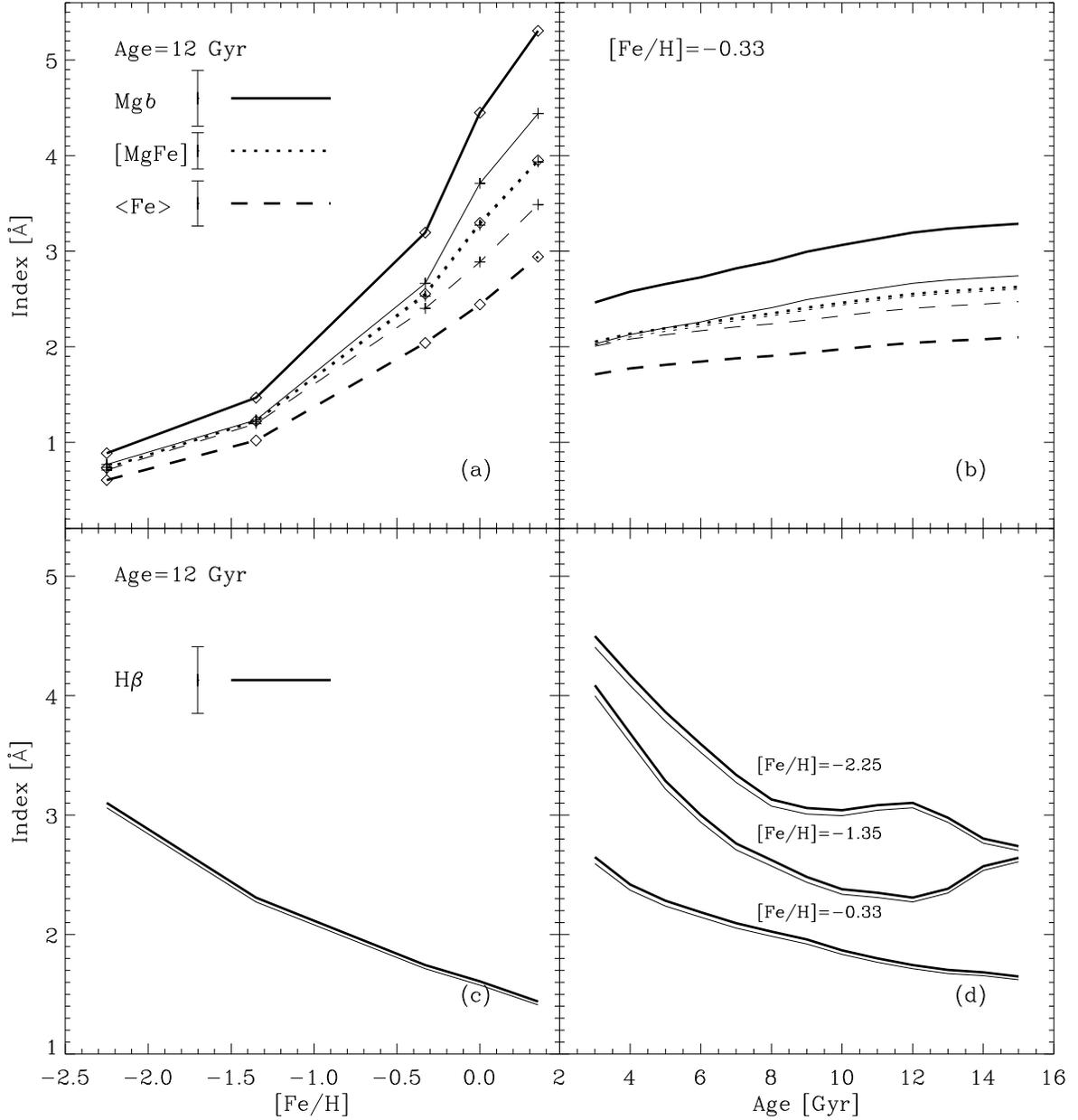}}
\caption{Dependence of \mgb, [MgFe], \meanFe\/ and \hb\/ on
  [Fe/H] and age for models with different [Mg/Fe] as predicted by
  \citet{TMB02b}. The thick lines and open diamonds stand for [Mg/Fe]
  $=+0.3$ whereas the thin lines and plus symbols represent [Mg/Fe]
  $=0.0$. The respective line-style and average measurement error for
  the different indices is given on the left hand side of panels (a)
  and (c). For the [MgFe] index the two model lines (for [Mg/Fe] $=0.0$
  and 0.3) overlap almost completely and therefore the index is almost
  independent of the Mg to Fe abundance ratios.}
           \label{fig:models}%
\end{figure*}

\citet{TMB02b} \citep[see also][]{TMB02a} provide new models which take
the stellar library biases into account and can predict line-strengths
for solar abundance as well as non-solar abundance ratios over a large
metallicity range ($-2.25 \le \mathrm{[Fe/H]} \le 0.35$). Since our
paper is among the first that make use of these new models for studying
GC spectra, we first explore systematically how the three parameters
age, metallicity and abundance ratio ([Mg/Fe]) affect the absorption
line-strengths of SSPs. For this purpose we plot in
Fig.~\ref{fig:models} the model predictions for the line-strengths of
the often used metallicity indicators \mgb, \meanFe\footnote{$\meanFe =
  ( \mathrm{Fe5270} + \mathrm{Fe5335})/2$\,, \citep{gon93}} \&
[MgFe]\footnote{$\mathrm{[MgFe]} = \sqrt{\mgb \times ( \mathrm{Fe5270}
    + \mathrm{Fe5335})/2}$\,, \citep{gon93}} and the age sensitive
Balmer line \hb\/ as a function of metallicity, age and [Mg/Fe].

The effects of non-solar abundance ratios (thick lines) with respect to
solar ratio model predictions (thin lines) can be clearly seen for the
metallicity sensitive indices \mgb\/ and \meanFe. At a given age and
all metallicities (Fig.~\ref{fig:models}a), the \mgb\/ index is
predicted to be stronger for super-solar [Mg/Fe]. The difference
between solar ratio and non-solar ratio models increases with
increasing metallicity. This behaviour is reversed for the \meanFe\/
index \citep[see also][]{tra00a}. A similar effect can be observed when
we fix [Fe/H] but vary age (Fig.~\ref{fig:models}b).  Remarkably, the
[MgFe] index, the geometric mean of the \mgb\/ and \meanFe\/ indices,
does not show any significant dependence on abundance ratios, at least
not within the current framework of the models. This has the unique
advantage that [MgFe] can be used as an empirical mean metallicity
indicator with negligible dependence on the abundance ratios \citep[see
also][]{kun01}. Similarly, \hb\/ is hardly affected by [Mg/Fe] with a
small increase in line-strength for larger [Mg/Fe] ratios
(Fig.~\ref{fig:models}c\&d).

However, the \hb\/ index shows a noteworthy complication at low
metallicities ($\mathrm{[Fe/H]} \le -1.35$) and ages $>$8~Gyrs. Here
the models predict a non-monotonic decrease of the \hb\/ index with
increasing age (see Fig.~\ref{fig:models}d, models for $\mathrm{[Fe/H]}
= -2.25$ \& $-1.35$). Therefore, in this age and metallicity range a
given measurement of \hb\/ and [MgFe] indices does not correspond to
one unique age but can in fact be consistent with a range of ages. The
ambiguity arises between a genuine measurement of the turn-off
temperature by \hb\/ and the appearance of blue horizontal branch stars
in old, metal poor stellar systems which will start to increase the
\hb\/ index, mimicking a younger age \citep{lee00,MT00,beas02b}.  The
overall result is a ``crossing'' of iso-age lines at ages larger than
8~Gyr and low metallicities. In Section~\ref{sec:agemetal} where we
will determine our best age estimates, the above effect will be taken
into account.

\section{Results}
\label{sec:results}
We first present our results on the abundance ratios of Mg to Fe of GCs
in NGC\,3115 and compare these with the Lick/IDS observations
\citep{tra98} of GCs in our own galaxy and M\,31. In
Section~\ref{sec:agemetal} we present our best estimates of the ages
and metallicities of NGC\,3115 GCs, while we compare photometric and
spectroscopic metallicity estimates in Section~\ref{sec:photspecz}.

\subsection{Abundance ratios}
\begin{figure*}
\centering
\includegraphics{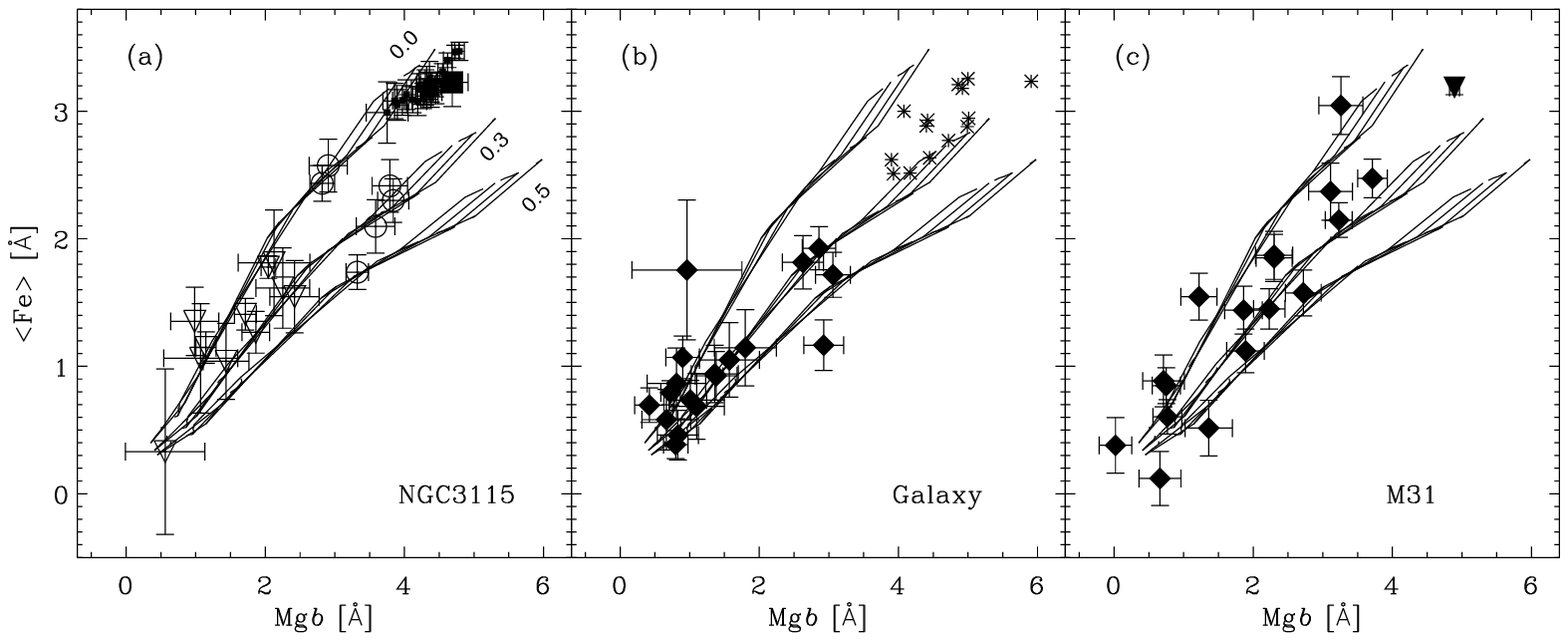}
\caption{\label{fig:abundance}Probing the [Mg/Fe] ratios of globular
  clusters in a \mgb\/ {\em vs}\/ \meanFe\/ diagram. Panel a): Our
  sample of globular clusters in NGC\,3115 is shown as open triangles
  (blue clusters) and open circles (red clusters). The large filled
  square represents the centre of NGC\,3115 taken from \citet{tra98}
  and the small filled squares represent the data of \citet{fis96}
  which cover radii up to 40\arcsec\/ along the major axis. Panel b):
  Milky Way globular clusters observed with the Lick/IDS
  instrumentation. Overplotted as stars are early-type galaxies in the
  Fornax cluster with central velocity dispersion $\sigma \ge
  75$~\kms\/ \citep{kun00}. Panel c): Globular clusters in M\,31
  observed with the Lick/IDS instrumentation. The filled triangle
  represents the centre of M\,31 taken from \citet{tra98}.  Overplotted
  in all panels are models by \citet{TMB02b} with abundance ratios of
  [Mg/Fe]=0.0., 0.3, 0.5 as indicated in the left panel. The models
  span a range in age (3--12~Gyr) and metallicity ($\mathrm{[Fe/H]} =
  -2.35~\mathrm{to}~ +0.3$).}
\end{figure*}

In Fig.~\ref{fig:abundance}, we show a diagram of \mgb\/ {\em vs}\,
\meanFe\/ index. We plot our sample of GCs separated by $(V-I)$ colour
(see Fig.~\ref{fig:colors}) where red and blue clusters are shown as
open circles and open triangles respectively.  Overplotted are the
model predictions by \citet{TMB02b} for [Mg/Fe] $= 0.0, 0.3, 0.5$, age
$= 3, 5, 8, 12$~Gyr and [Fe/H]~$= -2.25, -1.35, -0.33, 0.0, +0.35$. In
this diagram the effects of age and metallicity are almost completely
degenerate and the sensitivity towards abundance ratios is maximised.
The models predict a significant change in line-strength for varying
abundance ratios at large metallicities/ages (i.e. large \mgb\/ \&
\meanFe\/ line-strength) while at low metallicities the dependence is
smaller. Therefore, at a given error in line-strength, abundance ratios
are determined more accurately at larger metallicities/ages.

\begin{figure}
\centering
\resizebox{\hsize}{!}{\includegraphics{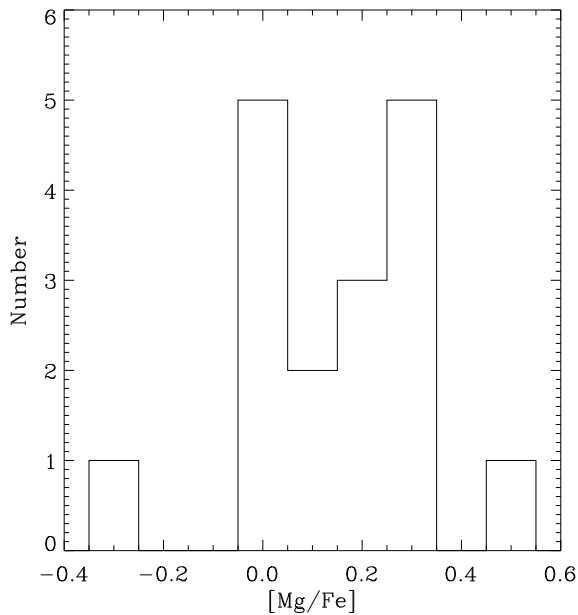}}
\caption{\label{fig:abundhist}Histogram of abundance ratios [Mg/Fe]
  for NGC\,3115 globular clusters. The abundance ratios were determined
  from Fig.~\ref{fig:abundance}a.}
\end{figure}

We find that for GCs in NGC\,3115 the abundance ratios vary from
roughly solar to about $\mathrm{[Mg/Fe]} \simeq +0.3$ (with a maximum
of $\mathrm{[Mg/Fe]} \simeq +0.5$, see Fig.~\ref{fig:abundance}a). In
our small sample there is no evidence for a trend with colour. In
particular, there are GCs with solar and super-solar abundance ratios
in both the blue and red GC populations. There is even some weak
evidence for a bimodal abundance ratio distribution with peaks at
[Mg/Fe] $\approx0.0$ and $\approx0.3$ (see Fig.~\ref{fig:abundhist}).
This tentative evidence needs to be confirmed with a larger sample of
GCs.

The measurements for the centre of NGC\,3115 itself \citep[large filled
square, data from][]{tra98} and the radial gradient \citep[small filled
squares, data from][]{fis96} along the major axis up to 40\arcsec\/ is
compatible with a model of $\mathrm{[Mg/Fe]} \approx 0.0$ at high
metallicity/age.  In Fig.s~\ref{fig:abundance}b and c we plot the
Lick/IDS observations of GCs in the Milky Way and M\,31 respectively
\citep[data from][]{tra98}.  While virtually all MW GCs are consistent
with [Mg/Fe]~$\simeq+0.3$, similar to large elliptical galaxies, we
find a range in [Mg/Fe] for the GCs in M\,31. The overall distribution
of the [Mg/Fe] ratios for GCs in M\,31 is similar to the one we find in
NGC\,3115. The average value of $\mathrm{[Mg/Fe]} = +0.3$ we find for
the MW GCs compares well with high resolution studies of individual
stars in MW GCs \citep[e.g.,][]{lee02}.

GCs with super-solar [Mg/Fe] ratios were previously found in other
nearby galaxies but without our quantitative accuracy. For example,
\citet{for01} attribute super-solar abundance ratios to 4 out of 10 GCs
in NGC\,1399. Using Mg and TiO features, \citet{lar02b} find a mean
[$\alpha$/Fe] of $+0.4$ for both metal-poor and metal-rich GCs in the
Sombrero galaxy.

\subsection{Age and metallicity}
\label{sec:agemetal}
In the previous section we were able to determine the abundance ratios
of GCs without knowing the age and metallicity since the latter
parameters are almost completely degenerate in a \mgb\/ {\em vs}\/
\meanFe\/ diagnostic diagram. However, our earlier discussion of the
model systematics shows that we need to take the abundance ratios into
account in order to estimate the age and metallicity of the GCs
\citep[see also e.g.,][]{tra00a, kun01}.

Principle age sensitive lines within our observed wavelength range are
the Balmer lines \hb, \hg, and \hd. For \hg\/ and \hd, the dependence
on $\alpha$--element to Fe ratio is yet unknown. \hb\/ is only
marginally sensitive to abundance ratio variations, at least in
comparison to our average observational error. To further minimise the
influence of abundance ratios, we employ as metallicity indicator
[MgFe], that also shows no significant [Mg/Fe] dependence (see
Fig.~\ref{fig:models}).  Within the accuracy of our data sample, an
\hb\/ {\em vs}\/ [MgFe] diagram can therefore be used to estimate the
ages and metallicities of our NGC\,3115 GCs without being significantly
affected by abundance ratios.

\begin{figure*}
  \centering \includegraphics[width=16cm]{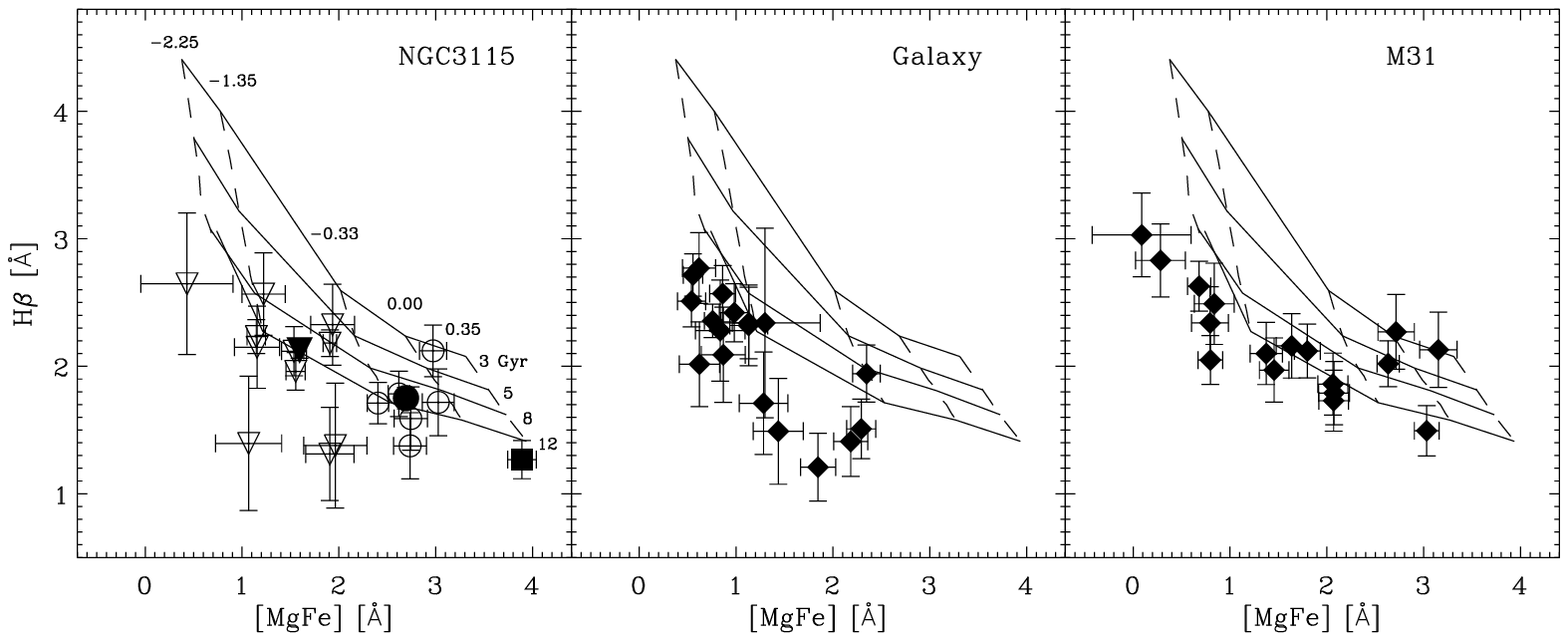}
  \centering \includegraphics[width=16cm]{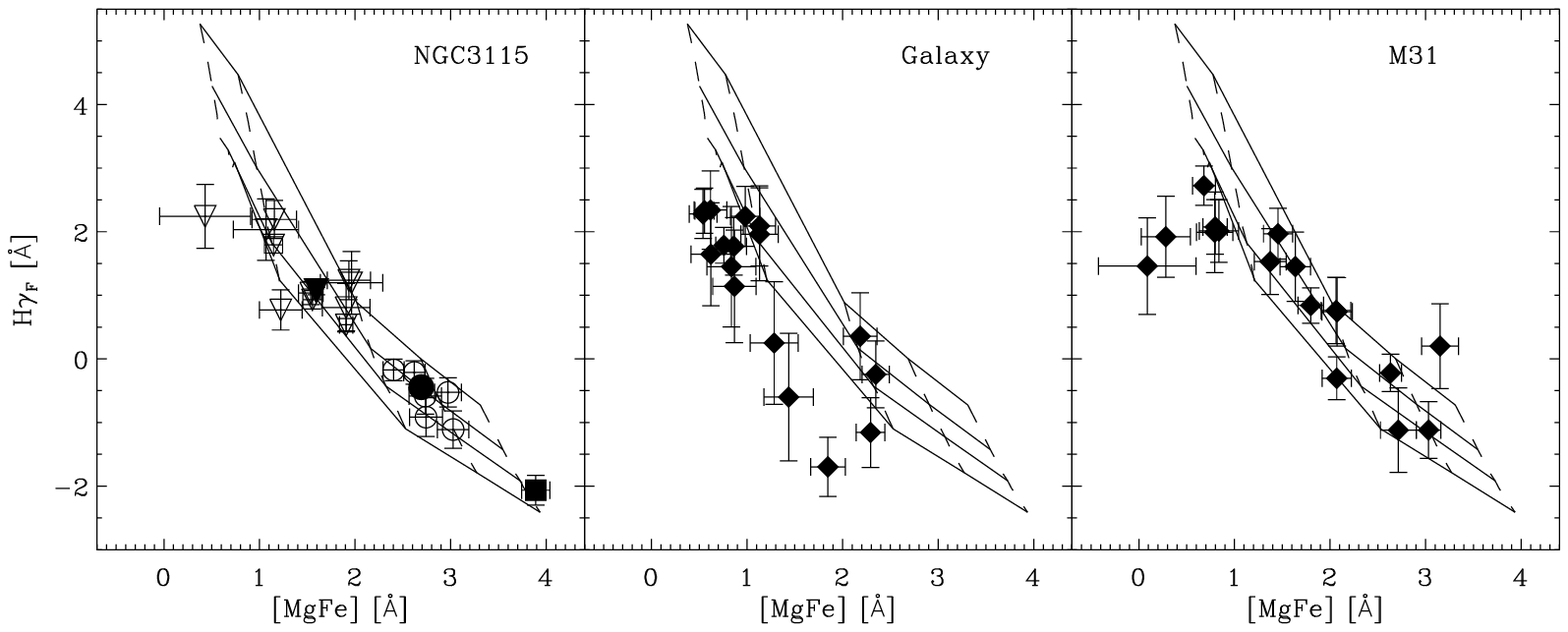}
  \centering \includegraphics[width=16cm]{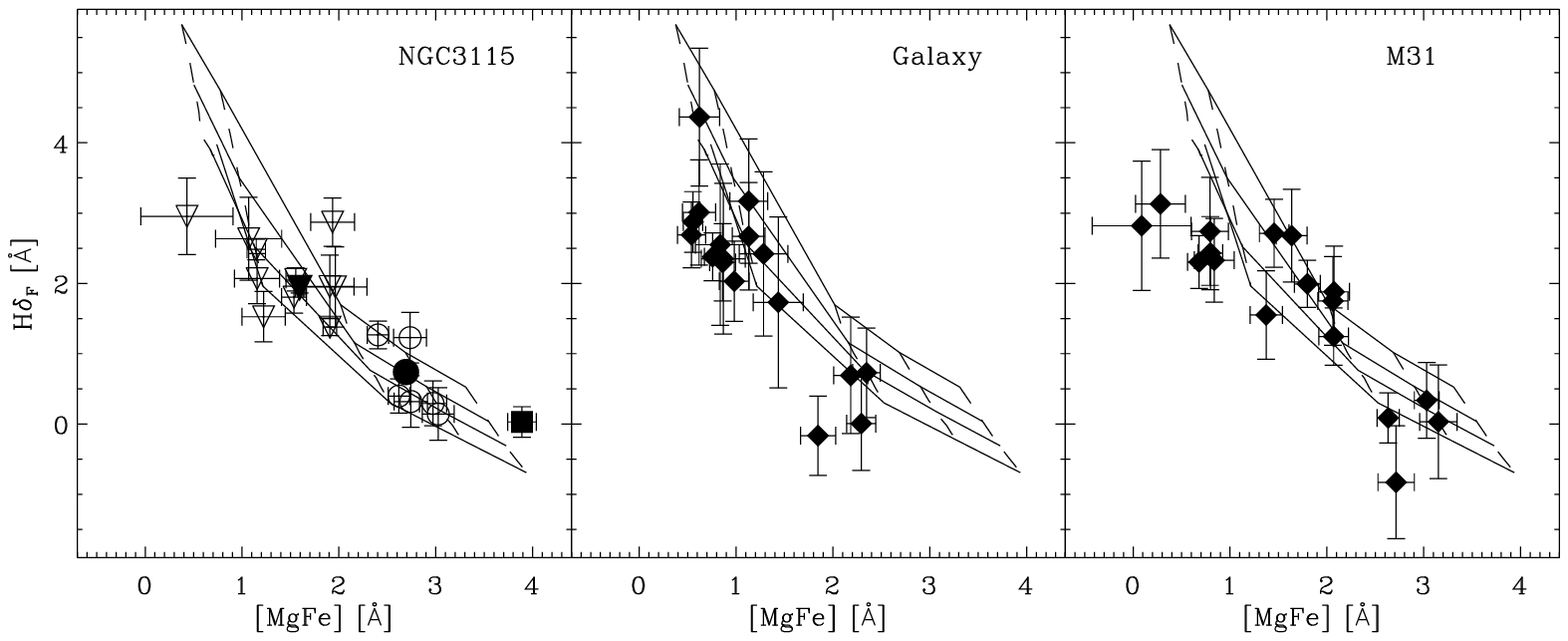}
\caption{Age and metallicity diagnostic diagrams using as metallicity
  indicator [MgFe] and as age indicator \hb, \hgf, and \hdf. \hb\ and
  [MgFe] are not significantly influenced by abundance ratios, for
  \hgf, and \hdf\/ the behaviour is unknown. Our sample of globular
  clusters in NGC\,3115 is shown in the left panels as open triangles
  (blue clusters) and open circles (red clusters). The error weighted
  means of the blue and red clusters are shown as filled symbols. The
  filled square in the left panels represents the centre of NGC\,3115
  \citet{tra98}. The middle and right panels show the Lick/IDS data for
  Milky Way MW and M\,31 globular clusters, respectively. Note, that
  for the MW globular cluster NGC\,2158 there are no \hgf\/ and \hdf\/
  index measurements available. Overplotted are solar--abundance SSP
  models by \citet{TMB02b} and Maraston (2002, in preparation) for
  metallicities $\mathrm{[Fe/H]} = -2.25, -1.35, -0.33, 0.00, 0.35$
  (dashed lines, left to right) and ages 3, 5, 8, and 12~Gyr (solid
  lines, top to bottom).}
           \label{fig:ages}%
\end{figure*}

In Fig.~\ref{fig:ages} we show diagrams of [MgFe] versus the three
Balmer lines for our sample of NGC\,3115 GCs (left panels) and the
respective data for GCs in the Milky Way and M\,31 from the Lick/IDS
observations (middle and right panels; the index values for \hb, \mgb\/
and \meanFe\/ were taken from \citet{tra98}; the higher order Balmer
lines of the Lick/IDS observations are presented in
Table~\ref{tab:galactic_lst} \& \ref{tab:m31_lst} in the Appendix).
Overplotted in Fig.~\ref{fig:ages} are solar--abundance ratio models by
\citet{TMB02b} and Maraston (2002, in preparation) for metallicities
[Fe/H] $= -2.25, -1.35, -0.33, 0.00, 0.35$ (dashed lines, left to
right) and ages 3, 5, 8 and 12~Gyr (solid lines, top to bottom).

We first caution that a direct comparison between models and our data
to derive {\em absolute}\/ ages and metallicities can be dangerous due
to possible systematic calibration errors. However, we estimate that
systematic observational errors are smaller than 0.1~\AA\/ for the
indices shown here and emphasize that relative comparisons within one
sample will be significant.

The observed \hb\/ values for NGC\,3115 GCs show a large spread with
respect to the model predictions. However, the data points which are
well below the model predictions are the ones with the largest errors.
Most of the well determined data points are close to the region of a
12~Gyr SSP model. Our [MgFe] measurements show that there is a clear
distinction in metallicity between blue (open triangles) and red
clusters (open circles), with the red clusters being more metal rich (a
more detailed analysis of the metallicity distribution is presented in
Section~\ref{sec:photspecz}).

Since only about half of our data points have small enough error bars
to be useful for an individual age/metallicity evaluation we also
calculate the error weighted mean of the blue and red clusters,
respectively. These average values (filled symbols in
Fig.~\ref{fig:ages}, left panels) give for the metal poor (blue)
population an age of 12.0($^{+1.5}_{-2.0}$)~Gyr and $\mathrm{[Fe/H]} =
-1.05(\pm0.09)$, while the metal rich (red) population has an estimated
age of 10.8($^{+1.7}_{-1.8}$)~Gyr and $\mathrm{[Fe/H]} =
-0.26(\pm0.05)$. The errors on the age and metallicity are quoted as
1$\sigma$ errors on the mean values.

We note in Section~\ref{sec:abrat} that for metallicities
$\mathrm{[Fe/H]} \le -1.35$ and an age larger than 8~Gyr the strength
of the \hb\/ and [MgFe] indices is not uniquely connected to one age
anymore. In fact there is a ``crossing'' of iso-age curves. For clarity
we do not plot iso-age lines for ages greater than 12~Gyr in
Fig.~\ref{fig:ages}, but this effect has been taken into account when
deriving the errors on our best age and metallicity estimates.

From the \hb\/ {\em vs}\/ [MgFe] diagram we conclude that within our
observational errors the two populations of GCs in NGC~3115 have the
same age of 11--12~Gyr (assuming the calibration of models and data is
accurate). The observed indices for the integrated light in the centre
of NGC\,3115 \citep[taken from][ and shown as filled square in
Fig.~\ref{fig:ages}]{tra98} are consistent with a luminosity weighted
age of $\approx$12~Gyr.

The Lick/IDS observations of MW GCs also show a significant number of
objects below the model predictions. We note that there are no
systematic observational offsets to be expected since the data was
taken with the original Lick/IDS system. We speculate that observations
of these GCs may be contaminated by fore/back-ground stars. New
spectroscopic observations of MW GCs \citep{puz02b} support this
hypothesis since the new observations do not show such low \hb\/
values. Consistent with recent age estimates from the resolved stellar
populations of MW GCs \citep[e.g.,][]{ros99,sal02} we do not find
evidence for clusters younger than $\approx$8~Gyr.

The Lick/IDS observations for M\,31 GCs show a relatively small scatter
close to a 12~Gyr model prediction, with only three, metal-rich
clusters showing evidence of a younger age. We note however, that for
metallicities $\mathrm{[Fe/H]} < -1.35$ the models seem to
systematically over-estimate the \hb\/ absorption strength (or
alternatively over-estimate the [MgFe] absorption strength).

In the next paragraphs we will present our measurements of the higher
order Balmer lines \hgf\/ and \hdf. We emphasize here that while these
indices can be measured with a higher precision than \hb, it is
currently unknown how these indices depend on abundance ratios.
Furthermore the absolute calibration of these indices has not yet been
investigated in as much detail as the \hb\/ index.

The distributions of \hgf\/ and \hdf\/ {\em vs}\/ [MgFe] are narrower
compared to \hb\/ {\em vs}\/ [MgFe] and mostly encompassed by the model
grid. The error weighted means for NGC\,3115 GCs indicate an age of
$\approx$7 and $\approx$5~Gyr for the blue and red clusters,
respectively. These average ages are substantially lower than what we
inferred from the \hb\/ {\em vs}\/ [MgFe] diagram. We note that our
observations of \hgf\/ and \hdf\/ for GCs in NGC\,3115 agree well with
the Lick/IDS observations of M\,31 and therefore we conclude that the
calibration of the models is not consistent between \hb\/ and the
higher order Balmer lines. Despite this absolute calibration problem we
find a good agreement in a relative sense between \hb, \hgf\/ and \hdf.
Therefore, at least to first order, we can say that the higher Balmer
indices are not significantly affected by abundance ratios in the
metallicity range probed by our data.

Comparing the distributions for MW and M\,31 GCs, we find that the MW
one is broader and offset towards smaller \hgf\/ and \hdf\/ absorption
consistent with the \hb\ measurements. We ascribe this to contaminated
observations for MW GCs (see above). The Lick/IDS data (particularly
the \hb\/ {\em vs}\/ [MgFe] diagram), suggest that perhaps $\sim$3
metal rich M\,31 GCs have younger ages (3--5~Gyr). Alternatively, one
could account for the strong \hb\/ absorption in these metal rich
clusters if the \hb\/ index is significantly influenced by an extended
blue horizontal branch in an otherwise old, metal rich stellar
population. \citet{MT00} show that this effect can play a role in {\em
  metal poor}\/ stellar populations, however, to date there is scarce
observational evidence for the existence of a {\em populous extended}\/
blue horizontal branch in metal rich clusters. \citet{rich97} detected
a blue horizontal branch in two metal rich MW GCs and \citet{fer01}
detected UV-excess stars in the core of 47~Tuc \citep[see
also][]{moe00}.

Few spectroscopic observations of GCs in early-type galaxies with
sufficient S/N to investigate these effects have been published.
\citet{for01} find that most of their 10 GCs in the giant elliptical
NGC\,1399 are old and compatible with a model age of 11~Gyr (using
models by Maraston 2002, in preparation). Only two GCs display such
large \hb\/ values that these have either a very young age of
$\sim2$~Gyr or are ``contaminated'' by a significant blue horizontal
branch population which causes large \hb\/ absorption. The authors
prefer the first interpretation. \citet{lar02a} present spectra of 14
GCs in the Sombrero Galaxy (NGC\,4594). Their analysis of the co-added
spectra of metal-poor and metal-rich GCs leads to age estimates between
10--15~Gyr. The majority (11 out of 14 GCs) of the spectroscopic sample
of \citet{schro02} of M\,81 GCs is compatible with old ages
\citep[using models by][]{wor94}. There is only one outlier with a very
high \hb\/ line strength.

In summary we conclude from our best calibrated diagram of \hb\/ {\em
  vs}\/ [MgFe] that the majority of our sample of GCs in NGC\,3115,
regardless of their metallicity, are consistent with an age of
$\approx$12~Gyr. Only one, metal rich cluster (Slitlet ID: 7) shows a
combination of \hb\/ and [MgFe] absorption strength which indicates an
age lower than 8~Gyr. The higher order Balmer lines indicate a narrow
distribution in age, with a hint of the metal rich clusters being
younger by $\approx$2~Gyr. The unknown dependence of the higher order
Balmer lines on abundance ratios makes this age difference highly
speculative. The absolute ages indicated by the higher order Balmer
lines are lower compared to the \hb\/ index.  We ascribe this age
difference to an inaccurate calibration of the higher order Balmer
lines in the current stellar population models. The Lick/IDS samples of
MW and M\,31 GCs also show old stellar populations; only $\sim$3 GCs in
M\,31 show tentative evidence of younger stars.

\subsection{Photometric versus spectroscopic metallicity estimates}
\label{sec:photspecz}
In this section we compare our spectroscopic metallicity estimates with
photometric methods and also investigate the general distribution of
metallicities. For this purpose we assume an average age of the GCs in
NGC\,3115 of 12~Gyr which is consistent with our findings in the
previous section.

\begin{figure*}
\resizebox{\hsize}{!}{\includegraphics{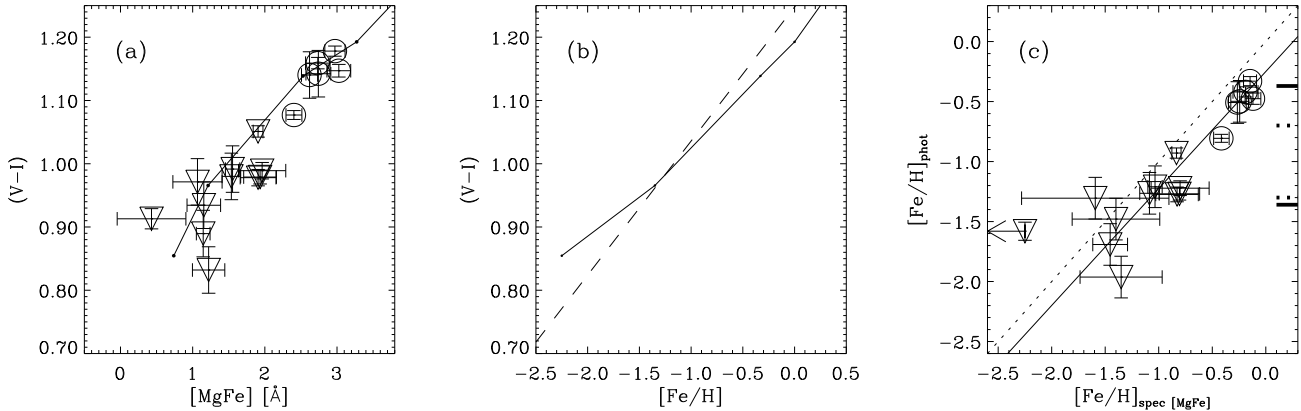}}
\caption{Comparison of photometric and spectroscopic metallicity
  estimates. In the left panel the [MgFe] index is plotted against
  $(V-I)$ colours (including the pseudo $(V-I)$ colours derived from
  the spectra). Overplotted as solid line is the prediction for a
  12~Gyr stellar population model by Maraston (2002, in preparation)
  and \citet{TMB02b}.  In the middle panel we show a comparison between
  an empirical calibration of $(V-I)$ colour against metallicity
  \citep[dashed line;][]{kundu98} and predictions from stellar
  population models (solid line, 12~Gyr; Maraston 2002, in
  preparation). In the right panel, photometric metallicities are
  calculated according to \citet{kundu98}, while spectroscopic
  metallicities are derived from the [MgFe] index in comparison to
  models of \citet{TMB02b} assuming a constant age of 12~Gyr. The data
  point at [Fe/H]$_\mathrm{spec~[MgFe]} = -2.25$ represents a globular
  cluster with line-strength weaker than the currently available model
  predictions and is therefore only an upper limit. The lines at the
  right hand side of the plot show the peak positions of the bimodal
  distributions of GCs observed by \citet[][ solid lines]{kundu98} and
  RGB stars observed by \citet[][ dotted lines]{els97}.}
           \label{fig:metal}%
\end{figure*}

Fig.~\ref{fig:metal}a shows the purely empirical relation between
$(V-I)$ colour and our mean metallicity indicator [MgFe]. There is a
tight relation over the observed parameter space. Overplotted as solid
line are model predictions by Maraston (2002, in preparation) and
\citet{TMB02b} for a constant age of 12\,Gyr, which is in excellent
agreement with our data.  We note that the model predictions for
colours do not include the effects of non-solar abundance ratios.

In order to convert the colours into metallicity estimates several
authors have derived linear conversion formulae based on observations
for MW GCs. For example, \citet{kundu98} conclude that [Fe/H] $= -5.89
+ 4.72 (V-I)$ is a good linear approximation. One can also use the
predictions of stellar population models (Maraston 2002, in
preparation) to predict the relation between $(V-I)$ colour and
metallicity [Fe/H]. A comparison of the empirical and synthetic
calibration (12~Gyr model) is shown in Fig.~\ref{fig:metal}b.

Overall, the agreement is acceptable, although there are significant
differences. Specifically at the low metalicity end the models predict
a shallower trend than the empirical relation. In order to stay
consistent with the \citet{kundu98} paper we use their relation to
convert $(V-I)$ colour to metallicity (see also
Table~\ref{tab:objects}). Furthermore we determine metallicity
estimates from our spectra by using the [MgFe] index in conjunction
with the model predictions by \citet{TMB02b} and assuming a constant
age of 12~Gyr. The comparison between photometric and spectroscopic
metallicity estimates is shown in Fig.~\ref{fig:metal}c.

We find a good linear relation between both methods. The best fitting
linear relation including the observational errors is
$\mathrm{[Fe/H]}_{\mathrm{phot}} = -0.26(\pm0.05) + 0.95(\pm0.08)
\times \mathrm{[Fe/H]}_{\mathrm{spec}}$ with a $\chi^2$ probability of
0.30. The systematic offset of approximately $\mathrm{[Fe/H]} = -0.26$
in the sense that the spectroscopic metallicity measurements are larger
is consistent with the difference between model predictions and
empirical calibration of the conversion formulae between colour and
[Fe/H] as shown in Fig.~\ref{fig:metal}b. The predicted non-linearity
of $(V-I)$ colour as function of metallicity below [Fe/H] $=-1.5$
cannot be tested since our data do not really cover this range.

In summary we confirm with our accurate spectroscopic observations that
the bimodal colour distribution seen in NGC\,3115 GCs is dominated by a
metallicity effect rather than by an age difference.  Furthermore, both
$(V-I)$ colour peaks do show a substantial spread in metallicity. We
conclude that in the metallicity range $-1.5 \le \mathrm{[Fe/H]} \le
0.0$ and in absence of young GCs, the $(V-I)$ colour is indeed a good
indicator for metallicity. We note that for metallicities below
$\mathrm{[Fe/H]} = -1.5$ this may not be the case anymore.

\section{Discussion}
\label{sec:discussion}
Before we discuss the implications of our results it needs to be noted
that our sample of GCs in NGC\,3115 is dominated by the bright end of
the luminosity function. Our faintest object (Slitlet ID: 17) which we
could use for the line-strength analysis is 0.9~mag brighter than the
peak of the GC luminosity function.  Furthermore, since this project
was only a pilot study we observed only a small number of clusters over
a limited region across NGC\,3115. Although our data can give clear
insights into the globular cluster system of NGC\,3115 it is not
necessarily a representative sample. The comparison with the Lick/IDS
observations of MW and M\,31 GCs needs to be viewed with caution since
these samples cover a larger range in absolute GC magnitude.

Using their HST $V$, $I$ photometry \citet{kundu98} develop the
following formation scenario for NGC\,3115. The blue, metal poor
clusters are formed with the halo/bulge component of the galaxy very
early on. Then about $4\pm3$~Gyr later an unequal mass, gas-rich merger
event forms the disk component and the associated red, metal rich
clusters. The authors point out that the spatial distribution of the
blue and red clusters are consistent with the halo/bulge and disk
components respectively. Furthermore, there is evidence from optical
imaging \citep{sil89}, that the disk is bluer and therefore perhaps
younger than the halo/bulge component, consistent with the above
outlined scenario.

In our spectroscopic study of the GCs in NGC\,3115 we find that the
clusters are consistent with being coeval at about 11--12~Gyr. There is
perhaps a weak hint of the red clusters being younger, but by no more
than $\sim$2~Gyr. This on its own would not rule out the scenario by
\citet{kundu98}, however our estimates of the abundance ratios show
that at least the population of red, metal rich clusters is not
homogeneous. For these objects we find a range in abundance ratios from
solar to about $\mathrm{[Mg/Fe]} \simeq 0.5$. In our small sample of
limited spatial extend we do not find any clear trends of the chemical
composition of GCs with the kinematics (see Figure~\ref{fig:kin}). The
relative velocities of most of the clusters in our sample are
consistent with rotation. A larger sample of more complete spatial
coverage is needed to establish possible trends between abundance
ratios and e.g., age, metallicity, spatial position and kinematics.

What we can however say is that the metal-rich clusters with {\em
  solar}\/ abundance ratios must have been made out of well mixed
material which incorporates the products of both SN\,II (the main
producer of alpha elements) and SN\,Ia (the main producer of Fe-peak
elements). Since SN\,Ia are somewhat delayed compared to SN\,II the
solar abundance ratio clusters must have formed after the initial star
burst in NGC\,3115. There are many possible scenarios to explain the
observed abundance ratio distribution, but it is hard to fit them into
a simple picture of only two distinct formation events which create the
red and blue globular cluster sub-populations \citep[see
also][]{beas02a}.

One scenario which we would like to put forward for further discussion
is the following. The metal poor (blue), non-solar abundance ratio
clusters are associated with the halo formation, and the metal-rich
(red), non-solar abundance ratio clusters are formed together with the
bulge as was similarly proposed for the Milky Way by \citet{CLL90} and
\citet{WG92}.  The metal rich, solar ratio clusters are then formed
with the disk of this lenticular galaxy $\sim$1--2~Gyr after the
initial star-burst perhaps in connection with a merger. This scenario
would then require the disk to have also close to solar abundance
ratios, which can be tested by observations of the integrated light.
Furthermore, if this scenario is correct the spatial distribution and
kinematics of disk GCs will be distinct from the population of halo and
bulge GCs in NGC\,3115. Future, larger samples of NGC\,3115 GCs will be
very valuable to explore the connection between disk formation and
metal rich globular cluster formation.

More spectroscopic observations of GCs in nearby galaxies of various
types will be very valuable to improve our stellar population models
and learn more about the early star-formation epochs in early-type
galaxies. However, the mismatch of the models and some observed indices
demonstrates that we are also in urgent need for a new, high-quality
flux-calibrated spectral library in order to exclude simple
observational offsets. Only then can we make good progress with
improvements on the input physics of stellar population models and
their application to extragalactic objects.

\section{Conclusions}
\label{sec:conclusions}
We present new, accurate measurements of absorption line-strength
indices of 17 globular clusters (hereafter GCs) in the nearby S0 galaxy
NGC\,3115. Our objects span a range in colour so that the bimodal
$(V-I)$ colour distribution is well sampled.

A critical comparison with Lick/IDS data \citep{tra98} of GCs in M\,31
and the Milky Way (hereafter MW) is presented. The Lick/IDS
measurements of the H$\gamma_{A,F}$ and H$\delta_{A,F}$ indices for MW
and M\,31 GCs are presented for the first time in this paper. The data
are analysed with new stellar population models \citep{TMB02b} which
are able to predict line-strength not only as a function of age and
metallicity but also as a function of abundance ratio.  Specifically,
abundance ratio biases in the stellar library, which is an essential
ingredient to the model predictions, have been taken into account for
the first time.

Our main results are listed in the following:

\begin{itemize}
  
\item The GCs in NGC\,3115 show a range of abundance ratios as
  estimated by the strength of Mg and Fe lines. Specifically we find
  for both red and blue clusters solar as well as super-solar values
  (up to $\mathrm{[Mg/Fe]} \simeq +0.5$). Lick/IDS data of M\,31 GCs
  show a similar distribution while MW GCs are consistent with a
  constant value of $\mathrm{[Mg/Fe]} \simeq 0.3$.  The latter is in
  agreement with recent studies of the resolved stellar populations in
  MW GCs.
  
\item Our analysis of \hb, \hgf\/ and \hdf\/ {\em vs}\/ [MgFe]
  age/metallicity diagnostic diagrams shows that the red and blue GC
  populations are coeval within $\approx$2~Gyr. Our best calibrated
  diagram (\hb, [MgFe]) indicates a mean age of $11-12$~Gyr. However,
  the higher order Balmer lines, although confirming the similar ages
  of blue and red clusters, indicate younger absolute ages
  ($\approx$6~Gyr). We ascribe these younger age estimates to
  inaccurate model calibrations of the higher order Balmer lines.
  Evidence for the existence of young ($<5$~Gyr) GCs in the studied
  galaxies is scarce. Only one cluster in NGC\,3115 and perhaps $\sim$3
  clusters in M\,31 show a combination of Balmer and metal absorption
  strength which is consistent with such young ages. The strong \hb\/
  absorption can be alternatively explained if these clusters have old
  stellar populations with a populous extended horizontal branch.
  
\item We present a comparison of photometric and spectroscopic
  metallicity determinations and find a good linear relation in the
  metallicity range probed by our sample of NGC\,3115 clusters ($-1.5 <
  \mathrm{[Fe/H]} < 0.0$). The photometric estimates are systematically
  lower ($\simeq -0.26$) in comparison with our spectroscopic
  measurements. We note that our observations clearly show that each
  colour peak has a significant spread in metallicity rather than being
  consistent with a narrow distribution.
  
\item The existence of solar as well as elevated Mg-to-Fe ratios at a
  given metallicity for GCs in NGC\,3115 indicates that a simple
  scenario of two distinct star-formation episodes is not sufficient to
  explain the formation of this galaxy. Probably a realistic model
  needs to incorporate more than two distinct star-formation events.
  
\item We detect a clear signal of rotation in our sample of GCs
  independent of their metallicities.
  
\end{itemize}

Larger samples of high signal-to-noise spectra of GCs in nearby
galaxies are required to increase our knowledge of how these galaxies
formed. Particularly the measurement of abundance ratios in
extragalactic GCs is now possible and will deliver important new
constraints for galaxy formation models which are not accessible with
broad-band colours alone.

\begin{acknowledgements}
  Part of this work was supported by the Volkswagen Foundation
  (I/76\,520). We thank Dr. J. Heidt (Heidelberg) and the ESO Paranal
  staff for the efficient observations, the resulting data of which
  were the basis for this Paper. We are also very grateful to
  D.~Thomas, C.~Maraston and R.~Bender who provided their models prior
  to publication. We thank the referee B.~W. Carney for a quick and
  helpful referee report.
\end{acknowledgements}

\bibliographystyle{aa}
\bibliography{references}

\begin{thebibliography}{60}
\expandafter\ifx\csname natexlab\endcsname\relax\def\natexlab#1{#1}\fi

\bibitem[{{Ashman} \& {Zepf}(1992)}]{AZ92}
{Ashman}, K.~M. \& {Zepf}, S.~E. 1992, \apj, 384, 50

\bibitem[{{Ashman} \& {Zepf}(1998)}]{ash97}
---. 1998, {Globular cluster systems} (Cambridge University Press)

\bibitem[{{Beasley} {et~al.}(2002{\natexlab{a}}){Beasley}, {Baugh}, {Forbes},
  {Sharples}, \& {Frenk}}]{beas02a}
{Beasley}, M.~A., {Baugh}, C.~M., {Forbes}, D.~A., {Sharples}, R.~M., \&
  {Frenk}, C.~S. 2002{\natexlab{a}}, \mnras, 333, 383

\bibitem[{{Beasley} {et~al.}(2002{\natexlab{b}}){Beasley}, {Hoyle}, \&
  {Sharples}}]{beas02b}
{Beasley}, M.~A., {Hoyle}, F., \& {Sharples}, R.~M. 2002{\natexlab{b}}, MNRAS,
  {accepted}

\bibitem[{{Bekki}(1998)}]{bekk98}
{Bekki}, K. 1998, \apjl, 502, L133

\bibitem[{{Borges} {et~al.}(1995){Borges}, {Idiart}, {de Freitas Pacheco}, \&
  {Thevenin}}]{BIDT95}
{Borges}, A.~C., {Idiart}, T.~P., {de Freitas Pacheco}, J.~A., \& {Thevenin},
  F. 1995, \aj, 110, 2408

\bibitem[{{Bridges} {et~al.}(1997){Bridges}, {Ashman}, {Zepf}, {Carter},
  {Hanes}, {Sharples}, \& {Kavelaars}}]{bri97}
{Bridges}, T.~J., {Ashman}, K.~M., {Zepf}, S.~E., {et~al.} 1997, \mnras, 284,
  376

\bibitem[{{Burstein} {et~al.}(1984){Burstein}, {Faber}, {Gaskell}, \&
  {Krumm}}]{bur84}
{Burstein}, D., {Faber}, S.~M., {Gaskell}, C.~M., \& {Krumm}, N. 1984, \apj,
  287, 586

\bibitem[{{Capaccioli} {et~al.}(1993){Capaccioli}, {Cappellaro}, {Held}, \&
  {Vietri}}]{cap93}
{Capaccioli}, M., {Cappellaro}, E., {Held}, E.~V., \& {Vietri}, M. 1993, \aap,
  274, 69

\bibitem[{{Carney} {et~al.}(1990){Carney}, {Latham}, \& {Laird}}]{CLL90}
{Carney}, B.~W., {Latham}, D.~W., \& {Laird}, J.~B. 1990, \aj, 99, 572

\bibitem[{Davies {et~al.}(1993)Davies, Sadler, \& Peletier}]{DSP93}
Davies, R.~L., Sadler, E.~M., \& Peletier, R.~F. 1993, \mnras, 262, 650

\bibitem[{{Dressler}(1980)}]{dres80}
{Dressler}, A. 1980, \apj, 236, 351

\bibitem[{{Dressler} {et~al.}(1997){Dressler}, {Oemler}, {Couch}, {Smail},
  {Ellis}, {Barger}, {Butcher}, {Poggianti}, \& {Sharples}}]{dres97}
{Dressler}, A., {Oemler}, A.~J., {Couch}, W.~J., {et~al.} 1997, \apj, 490, 577

\bibitem[{{Edvardsson} {et~al.}(1993){Edvardsson}, {Andersen}, {Gustafsson},
  {Lambert}, {Nissen}, \& {Tomkin}}]{edv93}
{Edvardsson}, B., {Andersen}, J., {Gustafsson}, B., {et~al.} 1993, \aap, 275,
  101

\bibitem[{{Elson}(1997)}]{els97}
{Elson}, R.~A.~W. 1997, \mnras, 286, 771

\bibitem[{{Ferraro} {et~al.}(2001){Ferraro}, {D'Amico}, {Possenti}, {Mignani},
  \& {Paltrinieri}}]{fer01}
{Ferraro}, F.~R., {D'Amico}, N., {Possenti}, A., {Mignani}, R.~P., \&
  {Paltrinieri}, B. 2001, \apj, 561, 337

\bibitem[{{Fisher} {et~al.}(1996){Fisher}, {Franx}, \& {Illingworth}}]{fis96}
{Fisher}, D., {Franx}, M., \& {Illingworth}, G. 1996, \apj, 459, 110

\bibitem[{{Forbes} {et~al.}(2001){Forbes}, {Beasley}, {Brodie}, \&
  {Kissler-Patig}}]{for01}
{Forbes}, D.~A., {Beasley}, M.~A., {Brodie}, J.~P., \& {Kissler-Patig}, M.
  2001, \apjl, 563, L143

\bibitem[{{Forbes} {et~al.}(1997){Forbes}, {Brodie}, \& {Grillmair}}]{forb97}
{Forbes}, D.~A., {Brodie}, J.~P., \& {Grillmair}, C.~J. 1997, \aj, 113, 1652

\bibitem[{{Gonz{\' a}lez}(1993)}]{gon93}
{Gonz{\' a}lez}, J.~J. 1993, Ph.D.~Thesis

\bibitem[{{Hanes} \& {Harris}(1986)}]{han86}
{Hanes}, D.~A. \& {Harris}, W.~E. 1986, \apj, 304, 599

\bibitem[{{Horne}(1986)}]{hor86}
{Horne}, K. 1986, \pasp, 98, 609

\bibitem[{{Kavelaars}(1998)}]{kav98b}
{Kavelaars}, J.~J. 1998, \pasp, 110, 758

\bibitem[{{Kissler-Patig} {et~al.}(1997){Kissler-Patig}, {Richtler}, {Storm},
  \& {della Valle}}]{kis97}
{Kissler-Patig}, M., {Richtler}, T., {Storm}, J., \& {della Valle}, M. 1997,
  \aap, 327, 503

\bibitem[{{Kundu} \& {Whitmore}(1998)}]{kundu98}
{Kundu}, A. \& {Whitmore}, B.~C. 1998, \aj, 116, 2841

\bibitem[{{Kuntschner}(2000)}]{kun00}
{Kuntschner}, H. 2000, \mnras, 315, 184

\bibitem[{{Kuntschner} {et~al.}(2001){Kuntschner}, {Lucey}, {Smith}, {Hudson},
  \& {Davies}}]{kun01}
{Kuntschner}, H., {Lucey}, J.~R., {Smith}, R.~J., {Hudson}, M.~J., \& {Davies},
  R.~L. 2001, \mnras, 323, 615

\bibitem[{{Larsen} \& {Brodie}(2002)}]{lar02a}
{Larsen}, S.~S. \& {Brodie}, J.~P. 2002, \aj, 123, 1488

\bibitem[{{Larsen} {et~al.}(2002){Larsen}, {Brodie}, {Beasley}, \&
  {Forbes}}]{lar02b}
{Larsen}, S.~S., {Brodie}, J.~P., {Beasley}, M.~A., \& {Forbes}, D.~A. 2002,
  \aj, 124, 828

\bibitem[{{Lee} {et~al.}(2000){Lee}, {Yoon}, \& {Lee}}]{lee00}
{Lee}, H., {Yoon}, S., \& {Lee}, Y. 2000, \aj, 120, 998

\bibitem[{{Lee} \& {Carney}(2002)}]{lee02}
{Lee}, J. \& {Carney}, B.~W. 2002, AJ accepted (astro--ph/0205393)

\bibitem[{Maraston \& Thomas(2000)}]{MT00}
Maraston, C. \& Thomas, D. 2000, \apj, 541, 126

\bibitem[{{McWilliam}(1997)}]{mcw97}
{McWilliam}, A. 1997, \araa, 35, 503

\bibitem[{{Michard} \& {Marchal}(1994)}]{MM94}
{Michard}, R. \& {Marchal}, J. 1994, \aaps, 105, 481

\bibitem[{{Moehler} {et~al.}(2000){Moehler}, {Landsman}, \& {Dorman}}]{moe00}
{Moehler}, S., {Landsman}, W.~B., \& {Dorman}, B. 2000, \aap, 361, 937

\bibitem[{{Peletier}(1989)}]{pel89}
{Peletier}, R.~F. 1989, Ph.D.~Thesis

\bibitem[{{Puzia} {et~al.}(2002{\natexlab{a}}){Puzia}, {Saglia},
  {Kissler-Patig}, {Maraston}, {Greggio}, {Renzini}, \& {Ortolani}}]{puz02b}
{Puzia}, T.~H., {Saglia}, R.~P., {Kissler-Patig}, M., {et~al.}
  2002{\natexlab{a}}, A\&A, submitted

\bibitem[{{Puzia} {et~al.}(2002{\natexlab{b}}){Puzia}, {Zepf}, {Kissler-Patig},
  {Hilker}, {Minniti}, \& {Goudfrooij}}]{puz02a}
{Puzia}, T.~H., {Zepf}, S.~E., {Kissler-Patig}, M., {et~al.}
  2002{\natexlab{b}}, \aap, 391, 453

\bibitem[{{Rich} {et~al.}(1997){Rich}, {Sosin}, {Djorgovski}, {Piotto}, {King},
  {Renzini}, {Phinney}, {Dorman}, {Liebert}, \& {Meylan}}]{rich97}
{Rich}, R.~M., {Sosin}, C., {Djorgovski}, S.~G., {et~al.} 1997, \apjl, 484, L25

\bibitem[{{Rosenberg} {et~al.}(1999){Rosenberg}, {Saviane}, {Piotto}, \&
  {Aparicio}}]{ros99}
{Rosenberg}, A., {Saviane}, I., {Piotto}, G., \& {Aparicio}, A. 1999, \aj, 118,
  2306

\bibitem[{{Salaris} \& {Weiss}(2002)}]{sal02}
{Salaris}, M. \& {Weiss}, A. 2002, \aap, 388, 492

\bibitem[{{Schlegel} {et~al.}(1998){Schlegel}, {Finkbeiner}, \&
  {Davis}}]{schle98}
{Schlegel}, D.~J., {Finkbeiner}, D.~P., \& {Davis}, M. 1998, \apj, 500, 525

\bibitem[{{Schroder} {et~al.}(2002){Schroder}, {Brodie}, {Kissler-Patig},
  {Huchra}, \& {Phillips}}]{schro02}
{Schroder}, L.~L., {Brodie}, J.~P., {Kissler-Patig}, M., {Huchra}, J.~P., \&
  {Phillips}, A.~C. 2002, \aj, 123, 2473

\bibitem[{{Sharples} {et~al.}(1998){Sharples}, {Zepf}, {Bridges}, {Hanes},
  {Carter}, {Ashman}, \& {Geisler}}]{sharp98}
{Sharples}, R.~M., {Zepf}, S.~E., {Bridges}, T.~J., {et~al.} 1998, \aj, 115,
  2337

\bibitem[{{Silva} {et~al.}(1989){Silva}, {Boroson}, {Thompson}, \&
  {Jedrzejewski}}]{sil89}
{Silva}, D.~R., {Boroson}, T.~A., {Thompson}, I.~B., \& {Jedrzejewski}, R.~I.
  1989, \aj, 98, 131

\bibitem[{{Smith} {et~al.}(2000){Smith}, {Lucey}, {Hudson}, {Schlegel}, \&
  {Davies}}]{smi00}
{Smith}, R.~J., {Lucey}, J.~R., {Hudson}, M.~J., {Schlegel}, D.~J., \&
  {Davies}, R.~L. 2000, \mnras, 313, 469

\bibitem[{Thomas {et~al.}(1999)Thomas, Greggio, \& Bender}]{TGB99}
Thomas, D., Greggio, L., \& Bender, R. 1999, \mnras, 302, 537

\bibitem[{{Thomas} {et~al.}(2002{\natexlab{a}}){Thomas}, {Maraston}, \&
  {Bender}}]{TMB02b}
{Thomas}, D., {Maraston}, C., \& {Bender}, R. 2002{\natexlab{a}}, MNRAS,
  submitted

\bibitem[{{Thomas} {et~al.}(2002{\natexlab{b}}){Thomas}, {Maraston}, \&
  {Bender}}]{TMB02a}
{Thomas}, D., {Maraston}, C., \& {Bender}, R. 2002{\natexlab{b}}, in Reviews in
  Modern Astronomy No.\,15, Astronomische Gesellschaft, (astro--ph/0202166)

\bibitem[{{Tinsley}(1979)}]{tin79}
{Tinsley}, B.~M. 1979, \apj, 229, 1046

\bibitem[{{Tonry} {et~al.}(2001){Tonry}, {Dressler}, {Blakeslee}, {Ajhar},
  {Fletcher}, {Luppino}, {Metzger}, \& {Moore}}]{ton01}
{Tonry}, J.~L., {Dressler}, A., {Blakeslee}, J.~P., {et~al.} 2001, \apj, 546,
  681

\bibitem[{{Trager} {et~al.}(2000){Trager}, {Faber}, {Worthey}, \& {Gonz{\'
  a}lez}}]{tra00a}
{Trager}, S.~C., {Faber}, S.~M., {Worthey}, G., \& {Gonz{\' a}lez}, J.~J.~.
  2000, \aj, 119, 1645

\bibitem[{{Trager} {et~al.}(1998){Trager}, {Worthey}, {Faber}, {Burstein}, \&
  {Gonzalez}}]{tra98}
{Trager}, S.~C., {Worthey}, G., {Faber}, S.~M., {Burstein}, D., \& {Gonzalez},
  J.~J. 1998, \apjs, 116, 1

\bibitem[{{van Dokkum}(2001)}]{vdok01}
{van Dokkum}, P.~G. 2001, \pasp, 113, 1420

\bibitem[{{Vazdekis}(1999)}]{vaz99}
{Vazdekis}, A. 1999, \apj, 513, 224

\bibitem[{{Worthey}(1994)}]{wor94}
{Worthey}, G. 1994, \apjs, 95, 107

\bibitem[{{Worthey}(1998)}]{wor98}
---. 1998, \pasp, 110, 888

\bibitem[{{Worthey} {et~al.}(1992){Worthey}, {Faber}, \& {Gonzalez}}]{WFG92}
{Worthey}, G., {Faber}, S.~M., \& {Gonzalez}, J.~J. 1992, \apj, 398, 69

\bibitem[{{Worthey} \& {Ottaviani}(1997)}]{worott97}
{Worthey}, G. \& {Ottaviani}, D.~L. 1997, \apjs, 111, 377

\bibitem[{{Wyse} \& {Gilmore}(1992)}]{WG92}
{Wyse}, R.~F.~G. \& {Gilmore}, G. 1992, \aj, 104, 144

\end{thebibliography}

\appendix
\section{Tables of line-strength measurements}
\subsection{Line-strength measurements for globular clusters in NGC\,3115}
Table~\ref{tab:ngc3115_lst1} \& \ref{tab:ngc3115_lst2} present the
line-strength measurements for globular clusters in NGC\,3115. The
spectra have been broadened to the Lick/IDS resolution prior to index
measurements. Small offsets have been made to match the Lick system as
closely as possible (see Section~\ref{sec:obs} for details). For a
summary of the definition of Lick/IDS indices see \citet{tra98}.

\begin{table*}
  \caption{Line-strength measurements for NGC\,3115 globular clusters }
  \label{tab:ngc3115_lst1}
  \begin{tabular}{crrrrrrrrrr}\hline
Slitlet ID & \hda & \hdf  & CN$_1$ & CN$_2$ &  Ca4227& G4300& \hga &  \hgf&  Fe4383& Ca4455 \\
           & [\AA]&[\AA]  &[mag]   &[mag]   &[\AA]   &[\AA] &[\AA] &[\AA] &[\AA]   &[\AA]  \\\hline
    3   & $-1.41$ &  0.32 &  0.049 &  0.076 &  0.95  & 4.64 & $-4.76$& $-0.92$ &  3.74  & 1.05   \\
        &  0.54   &  0.37 &  0.015 &  0.018 &  0.26  & 0.43 &  0.51& 0.31 &  0.59  & 0.29   \\
    5   &  1.44   &  1.81 &$-0.014$&  0.022 &  0.54  & 2.84 & $-0.92$& 1.04 &  2.03  & 0.40   \\
        &  0.34   &  0.23 &  0.010 &  0.012 &  0.18  & 0.30 &  0.32& 0.18 &  0.44  & 0.21   \\
    7   & $-1.00$ &  0.29 &  0.078 &  0.120 &  1.41  & 4.45 & $-4.60$& $-0.53$ &  3.81  & 0.67   \\
        &  0.46   &  0.32 &  0.013 &  0.015 &  0.21  & 0.35 &  0.38& 0.23 &  0.49  & 0.22   \\
    8   &  0.49   &  1.38 &  0.006 &  0.038 &  0.83  & 3.29 & $-1.92$& 0.54 &  2.36  & 0.56   \\
        &  0.18   &  0.12 &  0.005 &  0.006 &  0.09  & 0.15 &  0.16& 0.10 &  0.22  & 0.11   \\
    9   &  0.68   &  1.27 &  0.015 &  0.054 &  0.79  & 3.91 & $-3.15$& $-0.17$ &  3.09  & 0.73   \\
        &  0.28   &  0.20 &  0.008 &  0.009 &  0.14  & 0.23 &  0.27& 0.17 &  0.35  & 0.17   \\
   12   &  2.42   &  2.87 &  0.010 &  0.067 &  0.60  & 2.38 & $-0.39$& 1.24 &  2.53  & 0.77   \\
        &  0.52   &  0.34 &  0.016 &  0.019 &  0.27  & 0.49 &  0.51& 0.30 &  0.71  & 0.35   \\
   13   &  2.02   &  1.95 &$-0.011$&  0.043 &  0.94  & 2.64 & $-1.78$& 1.20 &  2.62  & 0.04   \\
        &  0.81   &  0.57 &  0.023 &  0.028 &  0.42  & 0.72 &  0.81& 0.49 &  1.08  & 0.53   \\
   14b  &  0.75   &  1.23 &  0.012 &  0.035 &  1.01  & 3.93 & $-4.56$& $-0.58$ &  3.89  & 1.26   \\
        &  0.50   &  0.36 &  0.015 &  0.018 &  0.25  & 0.40 &  0.46& 0.28 &  0.60  & 0.28   \\
   15   &  2.54   &  1.96 &$-0.015$&  0.020 &  1.16  & 3.29 & $-1.72$& 0.81 &  3.00  & 0.12   \\
        &  0.64   &  0.45 &  0.019 &  0.023 &  0.34  & 0.56 &  0.64& 0.38 &  0.82  & 0.42   \\
   16   &  3.50   &  2.95 &$-0.063$&$-0.058$& $-0.21$  & 0.55 &  0.96& 2.24 &  2.07  & 0.60   \\
        &  0.84   &  0.55 &  0.025 &  0.030 &  0.45  & 0.83 &  0.78& 0.50 &  1.11  & 0.54   \\
   17   & $-0.27$ &  0.14 &  0.056 &  0.080 &  1.07  & 5.02 & $-4.59$& $-1.11$ &  3.53  & 0.47   \\
        &  0.54   &  0.37 &  0.014 &  0.018 &  0.24  & 0.40 &  0.48& 0.29 &  0.57  & 0.28   \\
   21   &  2.01   &  2.63 &$-0.048$&$-0.034$&  0.70  & 3.34 & $-0.17$& 2.03 &  1.85  & 0.17   \\
        &  0.85   &  0.59 &  0.024 &  0.029 &  0.43  & 0.79 &  0.79& 0.48 &  1.11  & 0.54   \\
   23   & $-1.79$ &  0.40 &  0.069 &  0.099 &  0.90  & 4.38 & $-3.97$& $-0.21$ &  4.77  & 1.13   \\
        &  0.37   &  0.24 &  0.010 &  0.012 &  0.16  & 0.28 &  0.31& 0.19 &  0.37  & 0.20   \\
   24   &  1.99   &  2.06 &$-0.008$&  0.012 &  0.83  & 2.87 & $-1.23$& 0.93 &  2.45  & 0.32   \\
        &  0.24   &  0.17 &  0.008 &  0.009 &  0.12  & 0.22 &  0.24  & 0.14 &  0.32  & 0.15   \\
   25a  &  1.41   &  1.53 &$-0.049$&$-0.041$&  0.32  & 2.66 & $-0.31$& 0.77 &  0.46  & 0.38  \\
        &  0.54   &  0.36 &  0.015 &  0.019 &  0.28  & 0.49 &  0.51  & 0.32 &  0.71  & 0.33  \\
   25b  &  2.33   &  2.48 &$-0.059$&$-0.034$&  0.49  & 2.33 &  0.30& 1.79 &  1.27  & 0.31   \\
        &  0.22   &  0.15 &  0.006 &  0.008 &  0.11  & 0.19 &  0.21& 0.13 &  0.31  & 0.15   \\
   26b  &  2.46   &  2.07 &$-0.064$&$-0.047$&  0.54  & 1.99 &  1.35& 2.19 &  1.33  & 0.03   \\
        &  0.53   &  0.35 &  0.016 &  0.019 &  0.28  & 0.49 &  0.49& 0.30 &  0.71  & 0.34   \\ \hline
  \end{tabular}
\medskip

\begin{minipage}{14.0cm}
  {\em Notes:}\/ Listed are line-strength measurements in the
  wavelength range 4000--4500~\AA\/ for globular clusters in NGC\,3115
  (see also Table~\ref{tab:objects}). The first line for each object
  shows the measured index values whereas the second line lists the
  associated errors. The unit for each index is given at the top in
  square brackets.
\end{minipage}

\end{table*}

\begin{table*}
  \caption{Lick/IDS line-strength measurements for NGC\,3115 globular
clusters }
  \label{tab:ngc3115_lst2}
  \begin{tabular}{crrrrrrrrrr}\hline
Slitlet ID &Fe4531& C$_2$4668 & \hb  &  Fe5015& Mg$_1$  &  Mg$_2$  &  \mgb&  Fe5270& Fe5335& Fe5406\\
           & [\AA]&[\AA]  &[\AA] & [\AA]  &[mag]&[mag]   &[\AA]  &[\AA]  &[\AA]  &[\AA]\\\hline
    3   &2.94  & 3.88  & 1.59 &  5.24  &0.068 & 0.216 &  3.59 &  2.30  &1.90  & 1.46\\
        &0.44  & 0.65  & 0.25 &  0.56  &0.006 & 0.007 &  0.28 &  0.28  &0.31  & 0.23\\
    5   &1.87  & 1.45  & 2.12 &  3.51  &0.038 & 0.105 &  1.86 &  1.49  &1.05  & 0.68\\
        &0.34  & 0.49  & 0.19 &  0.41  &0.005 & 0.005 &  0.20 &  0.22  &0.25  & 0.19\\
    7   &3.20  & 4.07  & 2.12 &  5.90  &0.064 & 0.196 &  3.84 &  2.03  &2.57  & 1.74\\
        &0.35  & 0.53  & 0.20 &  0.43  &0.005 & 0.006 &  0.22 &  0.24  &0.25  & 0.20\\
    8   &2.20  & 1.87  & 2.18 &  3.20  &0.025 & 0.110 &  2.04 &  1.83  &1.73  & 0.89\\
        &0.17  & 0.25  & 0.10 &  0.23  &0.002 & 0.003 &  0.10 &  0.12  &0.13  & 0.11\\
    9   &2.58  & 2.89  & 1.71 &  4.13  &0.050 & 0.160 &  3.32 &  1.97  &1.51  & 0.83\\
        &0.26  & 0.39  & 0.16 &  0.34  &0.004 & 0.004 &  0.16 &  0.18  &0.20  & 0.16\\
   12   &0.70  &$-0.29$  & 2.33 &  3.53  &0.025 & 0.118 &  2.42 &  1.88  &1.21  & 1.21\\
        &0.61  & 0.86  & 0.32 &  0.69  &0.008 & 0.010 &  0.35 &  0.37  &0.43  & 0.34\\
   13   &1.99  & 1.99  & 1.38 &  2.59  &0.014 & 0.107 &  2.12 &  2.73  &0.90  & 0.46\\
        &0.87  & 1.30  & 0.49 &  1.05  &0.011 & 0.013 &  0.52 &  0.54  &0.63  & 0.52\\
   14b  &3.26  & 1.87  & 1.38 &  3.75  &0.052 & 0.174 &  2.90 &  2.45  &2.69  & 1.22\\
        &0.43  & 0.67  & 0.26 &  0.57  &0.006 & 0.007 &  0.27 &  0.28  &0.31  & 0.26\\
   15   &1.68  &$-0.36$  & 1.31 &  3.85  &0.035 & 0.125 &  2.25 &  1.36  &1.87  & 0.47\\
        &0.66  & 1.00  & 0.37 &  0.80  &0.009 & 0.010 &  0.39 &  0.41  &0.48  & 0.41\\
   16   &1.64  &$-1.13$  & 2.65 &  1.62  &0.015 & 0.028 &  0.56 &  0.36  &0.30$^a$  &$-0.28$\\
        &0.92  & 1.48  & 0.56 &  1.18  &0.013 & 0.015 &  0.57 &  0.65  &1.12  & 0.57\\
   17   &3.18  & 3.47  & 1.72 &  4.80  &0.055 & 0.199 &  3.79 &  2.24  &2.59  & 1.19\\
        &0.43  & 0.66  & 0.26 &  0.52  &0.006 & 0.007 &  0.26 &  0.28  &0.30  & 0.25\\
   21   &0.66  & 0.26  & 1.39 &  1.48  &0.009 & 0.070 &  1.07 &  0.99  &1.13  & 1.11\\
        &0.92  & 1.37  & 0.53 &  1.11  &0.012 & 0.013 &  0.53 &  0.58  &0.63  & 0.53\\
   23   &3.21  & 2.36  & 1.78 &  4.31  &0.036 & 0.166 &  2.82 &  2.68  &2.19  & 1.65\\
        &0.30  & 0.45  & 0.18 &  0.36  &0.004 & 0.005 &  0.18 &  0.18  &0.21  & 0.16\\
   24   &2.01  & 0.46  & 1.96 &  3.04  &0.023 & 0.103 &  1.71 &  1.51  &1.31  & 1.06\\
        &0.25  & 0.39  & 0.15 &  0.32  &0.003 & 0.004 &  0.16 &  0.17  &0.19  & 0.15\\
   25a  &0.12  &$-0.86$& 2.57 &  0.97  &$-0.004$& 0.037& 1.44 &  1.29  &0.78  &$-0.33$\\
        &0.58  & 0.87  & 0.32 &  0.81  &0.008 & 0.009 &  0.33 &  0.38  &0.46  & 0.36\\
   25b  &1.60  & 0.61  & 2.23 &  2.66  &0.016 & 0.063 &  1.15 &  1.06  &1.23  & 0.50\\
        &0.23  & 0.33  & 0.13 &  0.30  &0.003 & 0.004 &  0.14 &  0.17  &0.18  & 0.14\\
   26b  &3.07  & 0.05  & 2.15 &  3.41  &0.027 & 0.088 &  0.99 &  1.52  &1.18  & 0.76\\
        &0.54  & 0.80  & 0.32 &  0.66  &0.007 & 0.009 &  0.35 &  0.37  &0.39  & 0.32\\\hline
  \end{tabular}
\medskip

\begin{minipage}{14.0cm}
  {\em Notes:}\/ Listed are line-strength measurements in the
  wavelength range 4500--5450~\AA\/ for globular clusters in NGC\,3115
  (see also Table~\ref{tab:objects}). The first line for each object
  shows the measured index values whereas the second line lists the
  associated errors. The unit for each index is given at the top in
  square brackets. \newline $^a$ For this globular cluster we measured
  a Fe5335 index of $-1.24(\pm0.75)$~\AA. This negative number is
  caused by a noisy section in the spectrum. By comparing this index
  with other Fe indices we established that a realistic value of the
  Fe5335 index for this globular cluster should be $+0.3$. To reflect
  the uncertainty in this procedure we increased the error by a factor
  of 1.5.
\end{minipage}

\end{table*}

\subsection{Lick/IDS observations of H$\gamma_{A,F}$ and H$\delta_{A,F}$
  indices for Milky Way and M\,31 globular clusters}
Since the higher order Balmer indices of Milky Way and M\,31 globular
clusters as measured in the Lick/IDS system have not yet been
published, we list the measurements of H$\gamma_{A,F}$ and
H$\delta_{A,F}$ in Tables~\ref{tab:galactic_lst} \& \ref{tab:m31_lst}.
The Lick observations are described in \citet{bur84}, names are
explained there \citep[see also][]{tra98}. M\,31 globular clusters were
observed with the standard aperture of the Lick/IDS system (1\farcs4
$\times$ 4\farcs0). The MW globular clusters were observed with a
long-slit of standard width that was raster-scanned on the sky to
create a square aperture of size 66\arcsec\/ $\times$ 66\arcsec\/ (only
for NGC\,6624 the aperture is 45\arcsec\/ $\times$ 60\arcsec). Raster
scans have the same spectral resolution as standard slit width scans.
For the definition of the higher order Balmer indices see
\citet{worott97}.

\begin{table}
  \caption{Lick/IDS line-strength measurements for Milky Way globular clusters}
  \label{tab:galactic_lst}
  \begin{tabular}{lrrrr}\hline
Name      & \hda  &  \hdf  &  \hga  &  \hgf \\
          & [\AA] &[\AA]   &[\AA]   &[\AA]  \\\hline
NGC\,5024 &  4.56 &   3.01 &   ...  &   2.34\\
          &  1.20 &   0.75 &        &   0.62\\
NGC\,5272 &  2.88 &   2.30 &   ...  &   1.77\\
          &  0.88 &   0.55 &        &   0.45\\
NGC\,5904 &  3.57 &   2.67 &   1.34 &   2.09\\
          &  1.22 &   0.76 &   0.92 &   0.63\\
NGC\,6171 &  2.18 &   1.73 &   ...  & $-0.60$\\
          &  1.95 &   1.22 &        &   1.00\\
NGC\,6205 &  2.68 &   2.38 &   1.04 &   1.79\\
          &  0.54 &   0.34 &   0.41 &   0.28\\
NGC\,6218 &  3.83 &   3.17 &   ...  &   1.96\\
          &  1.41 &   0.88 &        &   0.73\\
NGC\,6229 &  3.57 &   2.35 &   0.79 &   1.14\\
          &  1.72 &   1.07 &   1.29 &   0.88\\
NGC\,6341 &  3.65 &   2.87 &   1.72 &   2.33\\
          &  0.69 &   0.43 &   0.52 &   0.36\\
NGC\,6356 &  ...  &   0.01 &   ...  & $-1.16$\\
          &       &   0.66 &        &   0.55\\
NGC\,6624 &$-1.26$&   0.73 & $-4.70$& $-0.24$\\
          &  1.02 &   0.64 &   0.76 &   0.52\\
NGC\,6637 &$-1.37$& $-0.17$& $-4.23$& $-1.70$\\
          &  0.90 &   0.56 &   0.68 &   0.47\\
NGC\,6712 &  3.18 &   2.42 &   0.64 &   0.25\\
          &  1.87 &   1.17 &   1.40 &   0.96\\
NGC\,6838 &  0.66 &   0.69 & $-1.39$&   0.35\\
          &  1.32 &   0.83 &   0.99 &   0.68\\
NGC\,6981 &  3.24 &   2.55 &   ...  &   1.45\\
          &  1.84 &   1.15 &        &   0.95\\
NGC\,7006 &  3.45 &   4.37 &   ...  &   1.64\\
          &  1.57 &   0.98 &        &   0.81\\
NGC\,7078 &  3.37 &   2.69 &   1.85 &   2.28\\
          &  0.75 &   0.47 &   0.56 &   0.39\\
NGC\,7089 &  2.71 &   2.03 &   1.19 &   2.24\\
          &  0.92 &   0.57 &   0.69 &   0.47\\\hline
  \end{tabular}
\medskip

\begin{minipage}{6.5cm}
  {\em Notes:}\/ We list here the Lick/IDS measurements of the indices
  \hda, \hdf, \hga\/ and \hgf\/ for MW globular clusters. The Lick/IDS
  system is described in detail in \citet{tra98}. For the definition of
  the higher order Balmer indices see \citet{worott97}.  The first line
  for each object shows the measured index values whereas the second
  line lists the associated errors. The unit for each index is given at
  the top in square brackets. Three dots indicate where no index
  measurement is available.
\end{minipage}

\end{table}

%
%

\begin{table}
  \caption{Line-strength measurements for globular clusters in M\,31}
  \label{tab:m31_lst}
  \begin{tabular}{llrrrr}\hline
Name1   & Name2 &\hda  &  \hdf  &  \hga  &  \hgf \\
        &       &[\AA] &[\AA]   &[\AA]   &[\AA]  \\\hline
MII  & K1    &   1.29 &   1.99 & $-1.55$&   0.84\\
     &       &   0.54 &   0.34 &   0.40 &   0.28\\
MIV  & K219  &   3.36 &   2.30 &   2.72 &   2.72\\
     &       &   0.60 &   0.38 &   0.45 &   0.31\\
V4   & K33   &   3.86 &   2.82 &   0.68 &   1.46\\
     &       &   1.47 &   0.92 &   1.10 &   0.76\\
V12  & K76   &   1.81 &   1.55 &   0.12 &   1.53\\
     &       &   1.00 &   0.63 &   0.75 &   0.52\\
V23  & K119  &   2.38 &   2.68 & $-0.08$&   1.45\\
     &       &   1.06 &   0.66 &   0.79 &   0.54\\
V42  & K78   & $-0.19$&   1.24 &   ...  & $-0.30$\\
     &       &   0.65 &   0.41 &        &   0.34\\
V64  & K213  &   ...  &   1.88 &   ...  &   0.74\\
     &       &        &   0.65 &        &   0.54\\
V76  & K263  &   2.55 &   3.13 &   0.83 &   1.92\\
     &       &   1.24 &   0.77 &   0.93 &   0.64\\
V87  & K222  & $-2.58$& $-0.83$& $-3.95$& $-1.12$\\
     &       &   1.29 &   0.80 &   0.96 &   0.66\\
V95  & K229  &   3.39 &   2.71 &   0.93 &   1.97\\
     &       &   0.77 &   0.48 &   0.58 &   0.40\\
V99  & K286  &   3.44 &   2.33 &   0.64 &   2.01\\
     &       &   0.95 &   0.60 &   0.72 &   0.49\\
V100 & K217  & $-2.39$&   0.34 & $-5.56$& $-1.12$\\
     &       &   0.86 &   0.54 &   0.65 &   0.45\\
V116 & K244  & $-3.69$&   0.03 &   $-$  &   0.20\\
     &       &   1.29 &   0.81 &   $-$  &   0.67\\
V196 & K302  &   2.26 &   2.43 &   0.79 &   2.07\\
     &       &   0.83 &   0.52 &   0.62 &   0.43\\
V282 & K280  & $-1.36$&   0.09 & $-4.75$& $-0.22$\\
     &       &   0.57 &   0.36 &   0.43 &   0.29\\
V301 & K64   &   3.14 &   2.74 &   ...  &   1.99\\
     &       &   1.23 &   0.77 &        &   0.63\\
V101 & K272  &   ...  &   1.75 &   ...  &   0.76\\
     &       &        &   0.63 &        &   0.52\\\hline
  \end{tabular}
\medskip

\begin{minipage}{7.0cm}
  {\em Notes:}\/ We list here the Lick/IDS measurements of the indices
  \hda, \hdf, \hga\/ and \hgf\/ for globular clusters in M\,31. The
  Lick/IDS system is described in detail in \citet{tra98}. For the
  definition of the higher order Balmer indices see \citet{worott97}.
  The first line for each object shows the measured index values
  whereas the second line lists the associated errors. The unit for
  each index is given at the top in square brackets. Three dots
  indicate where no index measurement is available.
\end{minipage}

\end{table}

\clearpage

\end{document}